%% file: main2.tex
\documentclass[10pt]{article} 
\usepackage[preprint]{tmlr}

\input{math_commands.tex}

\usepackage{hyperref}
\usepackage{url}

\usepackage{mathtools}
\usepackage{wrapfig}
\usepackage{booktabs,makecell, multirow, tabularx}
\usepackage{xcolor}   
\usepackage{colortbl}

\usepackage{subcaption}

\title{Dual-branch Graph Domain Adaptation for Cross-scenario Multi-modal Emotion Recognition}

\author{\name Yuntao Shou \email shouyuntao@stu.xjtu.edu.cn \\
      \addr College of Computer and Mathematics \\ Central South University of Forestry and Technology\\
      \AND
      \name  Jun Zhou \email zhoujun@csuft.edu.cn \\
      \addr College of Computer and Mathematics \\ Central South University of Forestry and Technology\\
      \AND
      \name  Tao Meng \email mengtao@hnu.edu.cn\\
      \addr College of Computer and Mathematics \\ Central South University of Forestry and Technology\\
      \AND
      \name  Wei Ai \email aiwei@hnu.edu.cn\\
      \addr College of Computer and Mathematics \\ Central South University of Forestry and Technology\\
      \AND
      \name  Keqin Li \email lik@newpaltz.edu\\
      \addr Department of Computer Science \\
      State University of New York
     }

\def\month{MM}  
\def\year{YYYY} 
\def\openreview{\url{https://openreview.net/forum?id=XXXX}} 

\begin{document}

\maketitle

\begin{abstract}
Multimodal Emotion Recognition in Conversations (MERC) aims to predict speakers’ emotional states in multi-turn dialogues through text, audio, and visual cues. In real-world settings, conversation scenarios differ significantly in speakers, topics, styles, and noise levels. Existing MERC methods generally neglect these cross-scenario variations, limiting their ability to transfer models trained on a source domain to unseen target domains. To address this issue, we propose a Dual-branch Graph Domain Adaptation framework (DGDA) for multimodal emotion recognition under cross-scenario conditions. We first construct an emotion interaction graph to characterize complex emotional dependencies among utterances. A dual-branch encoder, consisting of a hypergraph neural network (HGNN) and a path neural network (PathNN), is then designed to explicitly model multivariate relationships and implicitly capture global dependencies. To enable out-of-domain generalization, a domain adversarial discriminator is introduced to learn invariant representations across domains. Furthermore, a regularization loss is incorporated to suppress the negative influence of noisy labels. To the best of our knowledge, DGDA is the first MERC framework that jointly addresses domain shift and label noise. Theoretical analysis provides tighter generalization bounds, and extensive experiments on IEMOCAP and MELD demonstrate that DGDA consistently outperforms strong baselines and better adapts to cross-scenario conversations. Our code is available at \href{https://github.com/Xudmm1239439/DGDA-Net}{https://github.com/Xudmm1239439/DGDA-Net}.
\end{abstract}

\section{Introduction}
Multimodal emotion recognition in conversations (MERC) \cite{shou2026comprehensive, shou2022conversational, ai2025revisiting, shou2024adversarial, shou2024low, meng2024deep, shou2025masked} aims to predict the emotional state of participants in a multi-round conversation through multimodal information (e.g., text, audio, and video) and has broad application prospects in dialogue generation \cite{ghosh2017affect, tu2024adaptive, zhang2024camel, tellamekala2023cold, wen2024personality, meng2024multi, shou2026graph, shou2024efficient}, social media analysis 
\cite{khare2020time, peng2024carat, xu2025multiple, huang2020challenges}, intelligent systems like smart homes and chatbots \cite{young2018augmenting, lu2025understanding, kang2025beyond, li2025revisiting, shou2025spegcl, ai2026paradigm}.

\begin{figure}[htbp]
	\centering
	\includegraphics[width=1\linewidth]{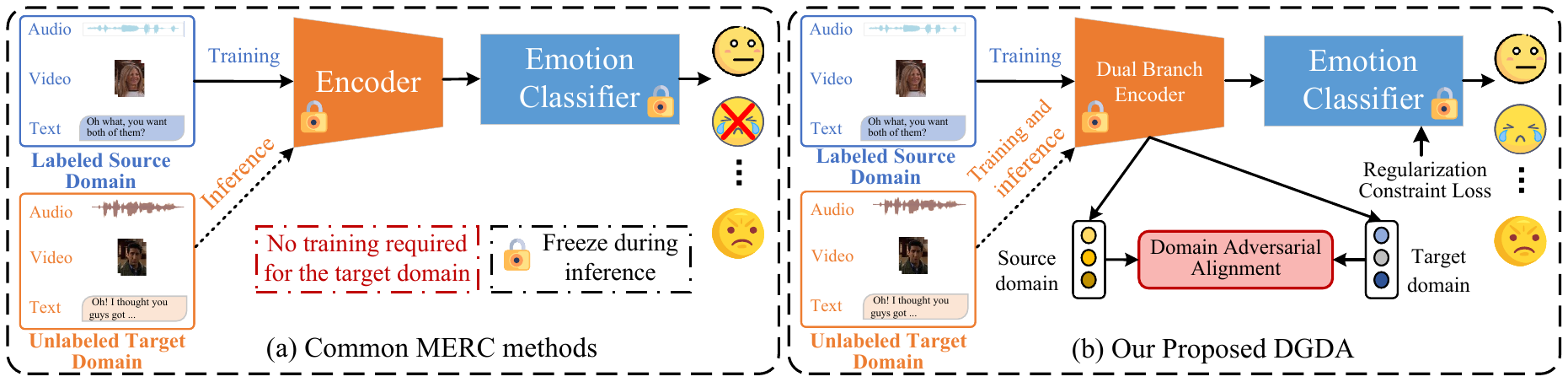}
	\caption{(a) Common MERC methods. A well-crafted encoder architecture is used to achieve multimodal emotion recognition without considering out-of-domain distribution differences. (b) Our Proposed Dual-branch Graph Domain Adaptation (DGDA) method. DGDA exploits a dual-branch encoder to explicitly and implicitly extract multimodal features, and constructs a domain adversarial alignment strategy and regularization loss to achieve out-of-domain distribution data generalization and resistance to noise label interference.}
	\label{fig:shouming}
\end{figure}

In the MERC task, researchers mainly focus on learning the emotional feature representation of in-domain data. As shown in Fig. \ref{fig:shouming} (a), a well-crafted encoder-classifier architecture is used to achieve multimodal emotion recognition without considering out-of-domain distribution differences. The mainstream MERC method mainly uses Transformers \cite{hazmoune2024using, lian2021ctnet, ma2023transformer, shou2025contrastive, shou2025revisiting} and graph neural networks (GNN) \cite{yang2024emotion, chen2023multivariate, shou2025dynamic, shou2025gsdnet, shou2023graphunet, shou2025graph} as the encoder to model contextual dependency information and speaker dependency. Although existing methods have achieved relatively good emotion recognition results, they ignore the impact of cross-domain distribution differences on emotion recognition performance. In other words, there may be significant differences in language styles, emotional expressions, and contextual environments in different domains, and these differences may lead to the limited generalization capabilities of the model. Furthermore, if some samples in the dataset are incorrectly labeled, e.g., the emotion of anger is incorrectly labeled as neutral, then the model may learn these wrong patterns during training. As a result, the model may misclassify the angry emotion as neutral in practical applications, thus affecting the accuracy and reliability of emotion recognition \cite{lian2023gcnet, wagner2023dawn, shou2025multimodal}.

To address the above problems, we propose a Dual-branch Graph Domain Adaptation (DGDA) for multi-modal emotion recognition in cross-scenario conversations, as shown in Fig. \ref{fig:shouming} (b). Specifically, to capture the discriminative features of emotion in multimodal utterances, we first construct an emotion interaction graph to model the complex emotional dependencies between utterances. Then, we design a hypergraph aggregation and path aggregation dual-branch graph encoder to explicitly and implicitly capture the dynamic changes in emotion between utterances and explore multivariate relationships, respectively. To address the problem of out-of-domain distribution differences, we introduce a domain adversarial classifier to improve the representation ability of invariant features in the source domain. In addition, we construct a regularization loss to prevent the model from memorizing noise and improve the model's ability to resist interference from noisy labels in source domains. Extensive experiments and evaluations demonstrate DGDA’s superiority.

The main contributions of this paper are summarized as follows:

\begin{itemize}
	\item To the best of our knowledge, we make the first attempt to simultaneously mitigate domain shift and noisy label interference problems in MERC scenarios, thereby enhancing usability in real-world scenarios.
	
	\item 
	We improve the expressiveness of domain-invariant features of the original graph by introducing a domain adversarial classifier and solving the problem of out-of-domain distribution differences.
	\item 
	We added a regularization constraint loss on the basis of the cross-entropy loss term to effectively suppress the model's over-learning of noisy labels and encourage the model to pay more attention to the real signals in the data.
	\item We provide theoretical proof to ensure that the designed DGDA is more precisely tailored for cross-scenario conversations. Extensive experiments conducted on the IEMOCAP and MELD datasets showed that DGDA is significantly better than existing baseline methods.
\end{itemize}

\section{Related work}
\subsection{Multimodal Emotion Recognition in Conversations}
\begin{figure*}
	\centering
	\includegraphics[width=1\linewidth]{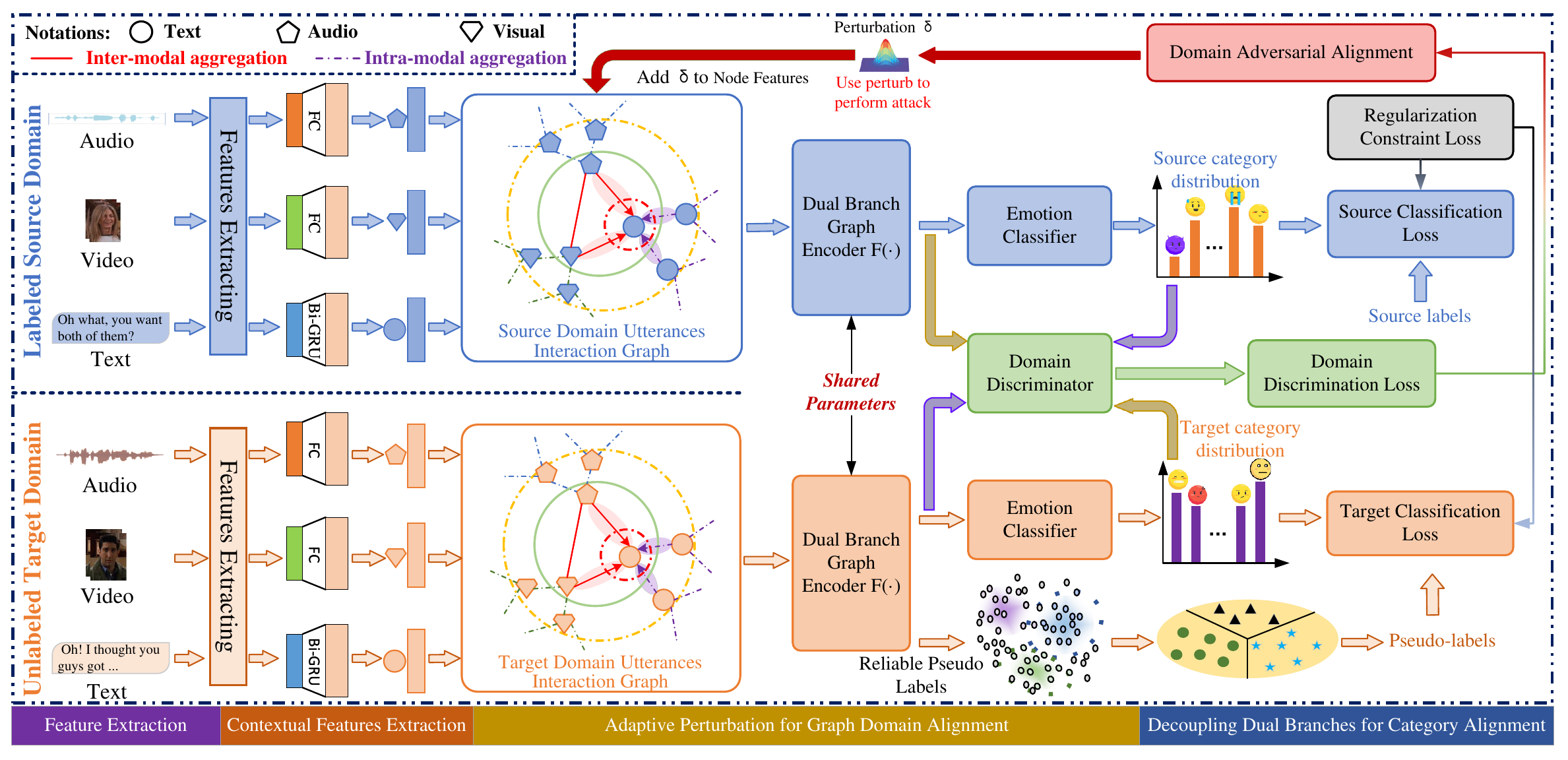}
	\caption{An overview of the proposed DGDA framework. The model operates on a labeled source domain and an unlabeled target domain. In both domains, audio, visual, and text features are first extracted and used to construct utterance-level interaction graphs. A dual-branch graph encoder encodes these graphs. For domain alignment, the source domain is adaptively perturbed by a learned noise $\delta$, while a domain discriminator promotes feature invariance across domains. Meanwhile, category-level alignment is enforced by coupling the dual-branch outputs. The final emotion classifier is trained using source labels and pseudo-labeled target samples.}
	\label{fig:archi}
\end{figure*}

Multimodal emotion recognition in conversations (MERC) has emerged as a key research area in artificial intelligence, especially at the intersection of natural language processing (NLP), computer vision (CV), and speech processing \cite{li2024cfn, liu2024eeg, tao2025multi, qin2025mental, shou2025graph, shou2025cilf}. Its objective is to infer human emotional states by jointly analyzing textual content, acoustic cues, and visual expressions \cite{sun2024muti, chen2024comprehensive, guo2025bridging, tang2025pose}. RNN-based MERC methods primarily focus on extracting contextual semantic information by modeling sequential dependencies within multimodal inputs through recurrent memory units \cite{majumder2019dialoguernn, huddar2021attention, ho2020multimodal}. Transformer-based methods leverage self-attention and multi-head attention mechanisms, often combined with pretrained language models, to capture long-range dependencies in conversations and achieve more effective multimodal fusion \cite{zhao2023tdfnet, ma2023transformer}. Meanwhile, GCN-based approaches utilize the structural flexibility of graph convolutional networks to model inter-utterance relations, multimodal associations, and latent interaction patterns within dialogues \cite{ren2021lr, yuan2023rba, ai2024gcn}. \textit{Despite their effectiveness within individual datasets, these methods generally overlook the challenges of cross-scenario multimodal emotion recognition and exhibit limited generalization when applied to out-of-domain conversational distributions.} Existing models often rely heavily on dataset-specific characteristics and struggle to maintain stable performance when domain shifts arise, such as variations in conversation styles, recording conditions, speaker demographics, or modality quality. This vulnerability leads to degraded robustness and restricts the deployment of MERC systems in real-world, heterogeneous environments. \textit{In contrast, as illustrated in Fig.~\ref{fig:archi}, our proposed DGDA framework explicitly addresses this limitation by introducing a domain adversarial classifier. This component encourages the model to learn domain-invariant feature representations through an adversarial optimization process, thereby mitigating domain discrepancies between the source and target distributions. By enhancing the extraction of shared, stable, and transferable multimodal features, DGDA significantly improves the model’s capability to generalize to out-of-domain conversational datasets and ensures more robust emotional understanding across diverse real-world scenarios.}

\subsection{Graph Domain Adaption}
Graph domain adaptation is a core issue in graph transfer learning \cite{qiu2020gcc, sun2022gppt, liu2024rethinking, zhang2025core, shou2024graph} and has received increasing attention in recent years, particularly in fields such as social networks and molecular biology \cite{you2023graph, chen2025smoothness, zhang2024collaborate}. Early studies mainly focused on transferring knowledge from a well-labeled source graph to an unlabeled target graph, aiming to learn discriminative representations for target graph nodes through label supervision from the source domain \cite{wu2020unsupervised, jin2024hgdl, dan2024tfgda}. These methods generally rely on propagating information along graph topology so that the target graph can inherit semantic cues and structural priors from the source graph. More recent research has further extended this paradigm to the graph-level setting, where multiple labeled source graphs must guide an unlabeled target graph \cite{yang2020heterogeneous, hu2024self}. In this scenario, the challenges go beyond simple node-level transfer; models must also handle semantic alignment, structural correspondence, and cross-graph knowledge integration at a holistic level \cite{yin2022deal}. Achieving such adaptation requires capturing similarities and discrepancies across heterogeneous graph distributions, reconciling different structural patterns, and transferring high-level semantic information. However, current graph-based domain adaptation methods face several fundamental limitations. Most models rely heavily on message-passing mechanisms that aggregate information from local neighborhoods. Although effective for learning localized patterns, such approaches struggle to capture high-order semantic dependencies, long-range relational structures, and global graph topology. As a result, they may fail to model complex structural variations between the source and target graphs, leading to insufficient domain alignment. In addition, existing methods typically assume that labels in the source domain are clean and reliable. In real-world scenarios, however, labeled graphs often contain noisy, ambiguous, or even contradictory annotations. Such label noise can propagate through the message-passing process, amplifying errors and degrading representation quality. The lack of explicit mechanisms to suppress noisy-label interference further limits the robustness and generalization performance of current approaches. \textit{Therefore, more advanced graph domain adaptation methods are needed to simultaneously capture global semantic structures and provide robustness against label noise, enabling more accurate and reliable cross-graph knowledge transfer.}


\section{METHODOLOGY}
\subsection{Task Definition}

In the task of Cross-Scenario Multimodal Emotion Recognition, we aim to build robust emotion recognition models that can generalize across diverse domains or scenarios, such as different datasets, environments, recording conditions, or speaker groups. Formally, we assume a set of speakers $S = \{s_1, s_2, \dots, s_M\}$ participating in emotionally rich conversations. Each conversation is composed of a sequence of utterances in chronological order, denoted as $U = \{u_1, u_2, \dots, u_N\}$, where $N$ is the total number of utterances. Each utterance $u_i$ is associated with a speaker $s_{p_i}$, defined through a speaker mapping function $p(\cdot): \{1, \dots, N\} \rightarrow \{1, \dots, M\}$. Furthermore, each utterance $u_{p_i}$ contains multimodal information, including textual modality $u^t_{p_i}$, visual modality $u^v_{p_i}$, and acoustic modality $u^a_{p_i}$. Unlike traditional emotion recognition tasks that assume training and testing data come from the same distribution, cross-scenario multimodal emotion recognition explicitly considers the domain shift between source and target scenarios. These shifts may arise due to variations in background, lighting, language usage, speaker identity, or even cultural differences. The objective is to predict the discrete emotion labels for each utterance in the target scenario, leveraging the multimodal information while ensuring robust generalization from the source to the target domain. This problem setting poses unique challenges, such as modality-specific noise, semantic gaps between scenarios, and inconsistent emotion distributions. Therefore, effective cross-scenario multimodal emotion recognition models must learn domain-invariant yet emotion-discriminative representations across modalities and scenarios.

\subsection{Multimodal Feature Extraction} 
We extract unimodal features at the utterance level as follows. Following our previous work \cite{ma2023transformer}, we introduce the RoBERTa Large model \cite{kim2021emoberta} for text feature extraction. The dimension of the final text feature representation is 1024. We use openSMILE \cite{eyben2010opensmile} to extract acoustic features. After feature extraction using openSMILE, we perform dimensionality reduction on the acoustic features through fully connected (FC) layers, reducing the feature dimension to 1582 for the IEMOCAP dataset and 300 for the MELD dataset. We use the DenseNet model \cite{huang2017densely} pre-trained on the Facial Expression Recognition Plus dataset for visual feature extraction. In the process of visual feature extraction, we use the output dimension of DenseNet as 342.

\subsection{Modal Feature Encoding}
For the MERC, the original dimensionality spaces of text, visual, and acoustic modalities are usually significantly different, which makes them not directly usable for graph construction or fusion. To address this problem, we design a shallow feature extractor that contains three independent encoders to map them into the same dimensional space. For text modality, we use a bidirectional gated recurrent unit (Bi-GRU) \cite{poria2017context} to capture the bidirectional dependencies of context. However, through empirical observation \cite{chen2023multivariate}, we find that using recurrent neural network modules to encode visual and acoustic modalities does not bring performance improvements. Therefore, we use a simple and efficient linear layer to convert them to the same dimensional space as the text modality. After feature extraction, we then input it into the subsequent emotion reasoning network.

\subsection{Dual Branch Encoder}

\textbf{Hypergraph neural network (HGNN) branch.} Traditional graph neural networks (GNNs) usually represent graph structures through nodes and edges, while hypergraph neural networks (HGNNs) further extend this framework by allowing a hyperedge to connect multiple nodes, thereby effectively modeling complex relationships and high-order dependencies. Specifically, given a sequence of utterances containing $N$ dialogue turns, we construct a hypergraph $H=(V,E,H)$, where $V$ is a set of nodes, $E$ is a set of hyperedges, and $H$ is an incidence matrix. Each node $v\in V$ corresponds to a unimodal utterance. Each hyperedge $e\in E$ encodes high-order dependencies between multimodal data and is assigned a weight $w(e)$. Each hyperedge $e\in E$ and each node $v\in V$ associated with $e$ is also assigned a weight $w_e(v)$. Incidence matrix $H\in \mathbb{R}^{|V| \times |E|}$ represents the relationship between nodes and hyperedges. If node $v$ is associated with hyperedge $e$, then $H_{ve} =1$, otherwise $H_{ve}=0$. Following the paradigm of previous work \cite{chen2023multivariate}, we adopted the node-hyperedge information interaction mechanism to achieve iterative update and fusion of features by alternately performing node convolution and hyperedge convolution. Mathematically,
\begin{equation}
	\mathbf{H}^{(l+1)} = \sigma \left( \mathbf{D}^{-1} \mathbf{H} \mathbf{W}_e \mathbf{B}^{-1} \hat{\mathbf{H}}^\top \mathbf{H}^{(l)} \right),
\end{equation}
where $\mathbf{H}^{(l)} = \{v_{i,(l)}^{x} | i \in [1, N], x \in \{t, a, v\}\} \in \mathbb{R}^{|\mathbf{H}| \times D}$ is the input at layer $l$, $v_{i}^{t}, v_{i}^{a}, v_{i}^{v}$ is the textual modality, acoustic modality and visual modality, respectively, $\mathbf{W}_e = \text{diag}(w(e_1), ..., w(e_{|{E}|}))$ is the hyperedge weight matrix, $\mathbf{B}$ and $\mathbf{D}$ are the hyperedge degree and node degree matrix, respectively.

\textbf{Path neural networks (PathNN) branch}. To effectively capture the global dependencies between nodes, we introduced PathNN \cite{michel2023path} into the model to make up for the lack of modeling capabilities of long-distance node relationships based on neighbor aggregation methods when dealing with complex graph structures. We first extract the feature representations of each node in the path based on the predefined paths in the graph, and then fuse the information of the nodes in the path into path representations through path aggregation operations. Finally, these path representations will be reversely injected into the start and end nodes associated with the path to achieve feedback of path information and enhancement of node representation. Suppose there is a set of paths $\mathcal{P}$, each path is represented as $p=(v_1, v_2,...,v _l)$, where $v_i$ is the $i$-th node in the path and $l$ is the path length. To further improve the expressiveness of node feature aggregation within a path and dynamically capture the importance of different nodes in path representation, we introduced an aggregation method based on the attention mechanism. We first calculate the attention score $\alpha_{vi}$ for each node in the path, defined as follows:
\begin{equation}
	\alpha_{v_i} = \frac{\exp \left( \text{LeakyReLU} \left( \mathbf{a}^\top [\mathbf{W} \mathbf{h}_{v_i} \parallel \mathbf{c}_p] \right) \right)}{\sum_{j=1}^{l} \exp \left( \text{LeakyReLU} \left( \mathbf{a}^\top [\mathbf{W} \mathbf{h}_{v_j} \parallel \mathbf{c}_p] \right) \right)},
\end{equation}
where $\mathbf{W}$ is a learnable matrix, $c_p$ is the path global context vector, $\mathbf{a}$ is a learnable attention weight vector, and $[\cdot\mid \mid \cdot]$ represents a vector concatenation operation. Next, the path representation $h_p$ is obtained by weighted summing of the node features in the path according to the attention weights:
\begin{equation}
	\mathbf{h}_p = \sum_{i=1}^{l} \alpha_{v_i} \cdot \mathbf{W} \mathbf{h}_{v_i}.
\end{equation}
Finally, the path information is passed back to the path-related nodes and updated as follows:
\begin{equation}
	\mathbf{h}_{v_i}^{(new)} = \mathbf{h}_{v_i}^{(old)} + \sum_{p \in \mathcal{P}(v_i)} \frac{1}{|\mathcal{P}(v_i)|} \mathbf{h}_p,
\end{equation}
where $\mathcal{P}(v_i)$ is the set of paths associated with node $v_i$.

\subsection{Adversarial Alignment for Domain Adaptation}

To alleviate the distribution difference between the source domain and the target domain in the multimodal graph semantic space and promote the consistency of the feature space, we introduced the adaptive perturbation and adversarial alignment mechanisms in both the HGNN branch and the PathNN branch. The core idea is to enhance the robustness of the model to domain changes by introducing learnable perturbations in the feature space, and dynamically optimizing the distribution of source domain and target domain features with the help of adversarial training to align them in the feature space. Specifically, for the original node feature $\mathbf{H}$ extracted by the HGNN branch and the PathNN branch, we first generate a perturbation vector for it through the adaptive perturbation generator and inject it into the original feature to obtain the perturbed feature representation:
\begin{equation}
	\begin{aligned}
		\widetilde{\mathbf{H}}^{HGNN} & = \mathbf{H}^{HGNN} + \delta^{HGNN} \cdot \mathbf{M}(\mathbf{H}^{HGNN}), \\ \widetilde{\mathbf{H}}^{PathNN} & = \mathbf{H}^{PathNN} + \delta^{PathNN} \cdot \mathbf{M}(\mathbf{H}^{PathNN}),
	\end{aligned}
\end{equation}
where $\delta$ is the perturbation intensity coefficient and $\mathbf{M}(\cdot)$ is the perturbation generation network, usually MLP. Next, we designed a domain discriminator to distinguish whether the features come from the source domain $\mathbf{H}_{src}$ or the target domain $\mathbf{H}_{tgt}$ as follows:
\begin{equation}
	\mathcal{L}_D = -\mathbb{E}_{\widetilde{\mathbf{H}}_{src}}[\log D(\widetilde{\mathbf{H}}_{src})] - \mathbb{E}_{\widetilde{\mathbf{H}}_{tgt}}[\log(1 - D(\widetilde{\mathbf{H}}_{tgt}))].
\end{equation}
The feature extractor (i.e., HGNN and PathNN branch) optimization objectives are as follows:
\begin{equation}
	\mathcal{L}_{adv} = -\mathbb{E}_{\widetilde{\mathbf{H}}_{tgt}}[\log D(\widetilde{\mathbf{H}}_{tgt})].
\end{equation}
During the training process, the domain discriminator and feature extractor are optimized alternately. The former attempts to correctly distinguish the source/target domain features, while the latter is continuously optimized by introducing perturbations and adversarial training so that the distributions of the two gradually converge.

\subsection{Branch Coupling for High-confidence Pseudo-label Generation}
\label{sec:them1}
In the cross-session multimodal emotion recognition task, the source domain data has reliable emotion labels, while the target domain has the problem of missing labels. Directly assigning pseudo labels to target samples for training often introduces noise due to low-confidence labels, affecting model performance. To this end, we propose branch coupling to generate high-confidence pseudo labels, aiming to make full use of the complementary features of the HGNN $p_{\theta}$ and the PathNN $q_{\phi}$ branch in graph semantic modeling and jointly optimize the quality of pseudo labels. We maximize the evidence lower bound (ELBO) of the log-likelihood with the source and target labels $y_s$ and $y_t$:
\begin{equation}
	\begin{aligned}
		\log p_{\theta}(y^{s}|G^{s},G^{t}) &= \log\int p_{\theta}(y^{s},{y}^{t}|G^{s},G^{t})\,d{y}^{t} = \log\int \frac{p_{\theta}(y^{s},{y}^{t}|G^{s},G^{t})}{q_{\phi}({y}^{t}|G^{t})}
		q_{\phi}({y}^{t}|G^{t})\,d{y}^{t} \\
		&\geq \int q_{\phi}({y}^{t}|G^{t})\log
		\frac{p_{\theta}(y^{s},{y}^{t}|G^{s},G^{t})}{q_{\phi}({y}^{t}|G^{t})}\,d{y}^{t} \\
		&\geq \mathbb{E}_{q_{\phi}({y}^{t}|G^{t})}\left[
		\log p_{\theta}(y^{s},{y}^{t}|G^{s},G^{t}) 
		- \log q_{\phi}({y}^{t}|G^{t})\right].
	\end{aligned}
	\label{eq:elbo}
\end{equation}

The traditional ELBO optimization strategy is mainly used to learn the approximate distribution $q$ of the target distribution $p$. However, the HGNN branch and the PathNN branch can model the potential label distribution of the samples respectively, one of which acts as the target distribution $p$, providing a relatively stable teacher signal, and the other as the approximate distribution $q$, which is as close to the target distribution as possible by maximizing the ELBO. To be specific, the ELBO in Eq. \ref{eq:elbo} can be equivalently written as:
\begin{equation}
	\begin{aligned}
		& \mathbb{E}_{q_{\phi}({y}^{t}|G^{t})}\left[\log p_{\theta}({y}^{t}|G^{s},G^{t},y^{s})p_{\theta}(y^{s}|G^{s})-\log q_{\phi}({y}^{t}|G^{t})\right] \\
		& =\mathbb{E}_{q_{\phi}({y}^{t}|G^{t})}\left[\log\frac{p_{\theta}({y}^{t}|G^{s},G^{t},y^{s})}{q_{\phi}({y}^{t}|G^{t})}\right]+\mathbb{E}_{q_{\phi}({y}^{t}|G^{t})}[p_{\theta}(y^{s}|G^{s})]\\
		&=-K L(q_{\phi}(y^{t}|G^{t})\parallel p_{\theta}(y^{t}|G^{s},G^{t},y^{s}))+\mathbb{E}_{q_{\phi}({y}^{t}|G^{t})}[p_{\theta}(y^{s}|G^{s})].
	\end{aligned}
\end{equation}

When optimizing the HGNN branch, we use the distribution $q_\phi$ of the PathNN branch output as the target and calculate the loss of the HGNN branch to approximate this distribution. Conversely, when optimizing the PathNN branch, we fix the distribution $p_\theta$ of the HGNN branch and use it as the optimization target of the PathNN branch. Based on the above ideas, we define the optimization loss functions of the HGNN and PathNN branches when they are updated alternately as follows:
\begin{equation}
	\begin{aligned}
		\mathcal{L}_{1}=&\mathbb{E}_{p_{\theta}\left(\hat{y}_{i}^{t}|G^{t}\right)>\zeta}\left[\log q_{\phi}\left(y_i^s, \hat{y}_{i}^{t}\mid  G_{i}^{s}, G_{i}^{t}\right)- \log q_{\phi}\left(\hat{y}_i^t\mid G_i^t\right)\right] \\
		&-\mathbb{E}_{q_{\phi}(y^s,G^s)}\log p_{\theta}\left(y_{i}^{s}\mid G_{i}^{s}\right), \\
		\mathcal{L}_2=&\mathbb{E}_{p_{\phi}\left(\hat{y}_{i}^{t}|G^{t}\right)>\zeta}\left[\log q_{\theta}\left(y_i^s, \hat{y}_{i}^{t}\mid  G_{i}^{s}, G_{i}^{t}\right)- \log q_{\theta}\left(\hat{y}_i^t\mid G_i^t\right)\right] \\
		&-\mathbb{E}_{q_{\theta}(y^s,G^s)}\log p_{\phi}\left(y_{i}^{s}\mid G_{i}^{s}\right),
	\end{aligned}
\end{equation}
where $\hat{y}_{i}^{t}$ is the target pseudo-label filtered by the HGNN or PathNN branch. It should be noted that we introduced a confidence threshold $\zeta$ to ensure that only those samples with higher confidence and more reliable prediction results can participate in the subsequent optimization process.

\textbf{Theorem 1.} For a deviation measure based on the label function $\max \limits_{G_{1},G_{2}}\frac{|\hat{h}_{D}(G_{1})-\hat{h}_{D}(G_{2})|}{\eta(G_{1},G_{2})}=C_{h}\le C_{f}C_{g}(D\in\{S,T\})$, let $H:=\{h:G\rightarrow Y\}$ denote a set of bounded real-valued functions that map from feature space $G$ to label space $Y$, the samples closest to the source domain distribution are selected in the target domain, the empirical risk $\hat{\epsilon}_{T}$ in the target domain can be significantly reduced:
\begin{equation}
	\begin{aligned}
		&\epsilon\tau(h,\hat{h}_{T})\le\frac{N_{T}^{\prime}}{N_{S}+N_{T}^{\prime}}\hat{\epsilon}_{T}(h,\hat{h}_{T}) +\frac{N_{S}}{N_{S}+N_{T}^{\prime}}\Big(\hat{\epsilon}_{S}(h,\hat{h}_{S})+\sqrt{\frac{4d}{N_{S}}\log(\frac{e N_{S}}{d})+\frac{1}{N_{S}}\log(\frac{1}{\delta})}\Big)
		\\& +\frac{N_{S}}{N_{S}+N_{T}^{\prime}}\Big(2C_{f}C_{g}W_{1}(\mathbb{P}_{S}(G),\mathbb{P}_{T}(G))+\omega\Big) \\ & \le\hat{\epsilon}_{S}(h,\hat{h})+\sqrt{\frac{4d}{N_{S}}\log(\frac{e N_{S}}{d})
			+\frac{1}{N_{S}}\log(\frac{1}{\delta})} + 2C_{f}C_{g}W_{1}(\mathbb{P}_{S}(G),\mathbb{P}_{T}(G))+\omega^{\prime},
	\end{aligned}
\end{equation}
where $\omega = \min_{\|g\|_{Lip} \leq C_g, \|f\|_{Lip} \leq C_f} \{ \epsilon_S(h, \hat{h}_S) + \epsilon_T(h, \hat{h}_S) \}$ and $\omega' = \min(|\epsilon_S(h, \hat{h}_S) - \epsilon_S(h, \hat{h}_T) \big|, \big| \epsilon_T(h, \hat{h}_S) - \epsilon_T(h - \hat{h}_T) \big|$.

\subsection{Model Training}

\label{them:23}

In cross-scenario emotion recognition, although the source domain data has labels, a certain proportion of labeling errors are inevitable in the real-world labeling process \cite{liu2020early}. Meanwhile, in the process of generating pseudo-labels in the target domain, inaccurate pseudo-labels are also inevitable. If noisy labels are used directly for model training, it is easy to cause the model to overfit the noise samples in the two branches of HGNN and PathNN. To alleviate this problem, we propose to introduce a regularization term in the cross-entropy loss. By introducing historical prediction results, the model can suppress the high-confidence fitting of labels in the later stage of training. Mathematically:
\begin{equation}
	\begin{aligned}
		\MoveEqLeft \mathcal{L}_{\text{CLS}} = -\frac{1}{N} \sum_{i=1}^{N} {y}_i \log p_{\theta}(y|x_i) 
		+ \lambda \cdot \left(\frac{1}{N} \sum_{i=1}^{N} \log \left( 1 - \langle p_{\theta}(y|x_i), \hat{p}_i \rangle \right)\right),
	\end{aligned}
\end{equation}
where $\lambda$ is the weight coefficient and $\hat{p}_i$ is the exponential moving average (EMA) of the model’s predicted probability for sample $x_i$ in the early training stage.

\textbf{Theorem 2.} Assume the input space is $\mathcal{X} \subseteq \mathbb{R}^d$ and the label space is $\mathcal{Y}=\{1,2,...,K\}$. The real data distribution is $\mathcal{D}$, and the observation distribution with noisy labels is $\tilde{D}$, where the upper limit of the noise ratio is $\eta \le 0.5$. In the early stage of training (the first $T_0$ steps), the model's prediction of clean samples satisfies $p^i_{\theta_i}(x) \approx y_i$, where $t \leq T_0$, and $(x,y)$ comes from the clean data subset ${\mathcal{D}}_{\text{clean}} \subset \tilde{\mathcal{D}}_2$, we then have:
\begin{equation}
	p_{\theta}(y_i|x_i) \approx \frac{\tilde{y}_i}{\tilde{y}_i + \lambda y_i (1 - p_{\theta}(y_i|x_i))}.
\end{equation}

\textbf{Theorem 3.} Assume that the model complexity is characterized by Rademacher complexity $\mathfrak{R}_n(\mathcal{F})$ \cite{yin2019rademacher}. For any $\delta >0$, the generalization error upper bound of $\mathcal{L}_{\text{CLS}}$ satisfies with probability $1 - \delta$:
\begin{equation}
	\begin{aligned}
		\MoveEqLeft\text{GenError}_{\mathcal{L}_{\text{CLS}}} \leq \text{GenError}_{\mathcal{L}_{\text{CE}}} \leq \frac{2 \mathfrak{R}_n(\mathcal{F})}{\sqrt{\lambda}} + \sqrt{\frac{\log(1/\delta)}{2n}} 
		+ O\left(\frac{\eta + \epsilon}{\mu}\right).
	\end{aligned}
\end{equation}

\section{DETAILED PROOFS}

\subsection{Proof of Theorem 1}
\label{sec:Theorem 1}

Intuitively, by combining training samples from the target and source domains, the class distributions between the two domains can be effectively aligned. Here, we provide a theoretical analysis to support this intuition. Specifically, we prove that after introducing the class distribution alignment module, the empirical risk lower bound in the target domain can be significantly reduced compared to models without this module. Before presenting our results, we first introduce a lemma \cite{redko2017theoretical, shen2018wasserstein, wang2024degree}, which is used in our proof:

\noindent\textbf{Lemma 1.} \emph{Let the learned discriminator $g$ be $C_{g}$-Lipschitz continuous, where the Lipschitz norm is defined as $||g||_{\mathrm{Lip}}=\max \limits_{Z_{1},Z_{2}}\frac{|g(Z_{1})-g(Z_{2})|}{\rho(Z_{1},Z_{2})}$, $\rho$ is a Euclidean distance function, and $H:=\{g:Z \rightarrow Y\}$ denote a set of bounded real-valued functions defined on the input space $Z$ and mapped to the output space $Y$. Assume that the pseudo-dimension of this set of functions is d, i.e., $Pdim(H)=d$. A bound on the relationship between empirical risk and true risk holds for the discriminator $g\in H$ with probability at least $1-\delta$:
	\begin{equation}
		\begin{aligned}
			\epsilon_{T}(h,\hat{h})\le&\hat{\epsilon}_{S}(h,\hat{h})+\sqrt{\frac{4d}{N_{S}}\log(\frac{e N_{S}}{d})+\frac{1}{N_{S}}\log(\frac{1}{\delta})}+2C_{f}C_{g}W_{1}(\mathbb{P}_{S}(G),\mathbb{P}_{T}(G))+\omega_{T}(G),
		\end{aligned}
	\end{equation}
	where $\omega=\mathrm{min}_{||{g}||_{L i p}\le C_{g}}\{\epsilon_{S}(g,\hat{g})+\epsilon_{T}(g,\hat{g})\}$ denotes the model discriminative ability, and the Wasserstein distance is defined as \cite{villani2009optimal}:
	\begin{equation}
		W_{1}(\mathbb{P},\mathbb{Q})=\operatorname*{sup}_{||g||L i p\le1}\left\{\mathbb{E}_{\mathbb{P}_{S}(Z)}g(Z)-\mathbb{E}_{\mathbb{P}_{T}(Z)}g(Z)\right\}.
\end{equation}}

Now, we present our theoretical results in the following theorem, as well as its proof.

\textbf{Theorem 1.} For a deviation measure based on the label function $\max \limits_{G_{1},G_{2}}\frac{|\hat{h}_{D}(G_{1})-\hat{h}_{D}(G_{2})|}{\eta(G_{1},G_{2})}=C_{h}\le C_{f}C_{g}(D\in\{S,T\})$, let $H:=\{h:G\rightarrow Y\}$ denote a set of bounded real-valued functions that map from feature space $G$ to label space $Y$, under the assumptions of Lemma 1 and the following assumptions:

And we have
\begin{enumerate} 
	\item Assume a small number of pseudo-labeled independent and identically distributed samples $\{(G_{n},Y_{n})\}_{n=1}^{N_{T}^{\prime}}$, where $N_{T}^{\prime}\ll N_{S}$ (the number of target domain samples is much smaller than source domain samples).
	\item Assume the source domain and the target domain have different label functions, satisfying $\hat{h}_{S}\ne\hat{h}_{T}$.  
	\item Assuming the samples closest to the source domain distribution are selected in the target domain, the empirical risk $\hat{\epsilon}_{T}$ in the target domain can be significantly reduced.    
	\begin{equation}
		\begin{aligned}
			\hat{\epsilon}_{T}\le\epsilon_{T}\le&\hat{\epsilon}_{S}(h,\hat{h})+\sqrt{\frac{4d}{N_{S}}\log(\frac{e N_{S}}{d})+\frac{1}{N_{S}}\log(\frac{1}{\delta})}
			+2C_{f}C_{g}W_{1}(\mathbb{P}_{S}(G),\mathbb{P}_{T}(G))+\omega^{\prime},
		\end{aligned}
	\end{equation}
	where $\omega^{\prime}=\operatorname*{min}_{||g|||_{L i p}\le C_{g},||f||_{L i p}\le C_{f}}\{\epsilon_{S}(h,\hat{h})+\epsilon_{T}(h,\hat{h})\}$, $\hat{\epsilon}_{T}$ is the empirical risk on the high confidence samples, $\epsilon_{S}$ is the empirical risk on the target domain.     
	\item Assume the pseudo-dimension of this set of functions is $Pdim(H)=d$. For any function $h\in H$, with probability of at least $1-\delta$, the following inequality holds:
	\begin{equation}
		\begin{aligned}
			\MoveEqLeft\epsilon_{T}(h,\hat{h}_{T}) 
			\leq\hat{\epsilon}_S(h,\hat{h}_S)+\sqrt{\frac{4d}{N_S}\log(\frac{eN_S}d)+\frac1{N_S}\log(\frac1\delta)} +2C_fC_gW_1(\mathbb{P}_S(G),\mathbb{P}_T(G))+\omega,
		\end{aligned}
	\end{equation}
	where $\omega=\operatorname*{min}_{||g|||_{L i p}\le C_{g},||f||_{L i p}\le C_{f}}\{\epsilon_{S}(h,\hat{h}_{S})+\epsilon_{T}(h,\hat{h}_{S})\}$.
\end{enumerate}

\textit{Proof.} We first introduce the following inequality to be used:
\begin{equation}
	\begin{aligned}
		\MoveEqLeft|\epsilon_{S}(h,\hat{h}_{S})-\epsilon_{T}(h,\hat{h}_{T})|
		=|\epsilon_{S}(h,\hat{h}_{S})-\epsilon_{S}(h,\hat{h}_{T})+\epsilon_{S}(h,\hat{h}_{T})-\epsilon_{T}(h,\hat{h}_{T})| \\
		&\leq|\epsilon_{S}(h,\hat{h}_{S})-\epsilon_{S}(h,\hat{h}_{T})|+|\epsilon_{S}(h,\hat{h}_{T})-\epsilon_{T}(h,\hat{h}_{T})| \\
		&\stackrel{(a)}{\leq}\left|\epsilon_{S}(h,\hat{h}_{S})-\epsilon_{S}(h,\hat{h}_{T})\right|+2C_{f}C_{g}W_{1}\left(\mathbb{P}_{S}(G),\mathbb{P}_{T}(G)\right),
	\end{aligned}
	\label{eq:4}
\end{equation}
where (a) results from \cite{shen2018wasserstein} Lemma 1 with the assumption $\max(||h||_{L i p},\max \limits_{G_{1},G_{2}}\frac{|\hat{h}_{D}(G_{1})-\hat{h}_{D}(G_{2})|}{\eta(G_{1},G_{2})})\le C_{f}C_{g},D\in\{S,T\}$. Similarly, we obtain:

\begin{equation}
	\begin{aligned}
		\MoveEqLeft|\epsilon_{S}(h,\hat{h}_{S})-\epsilon_{T}(h,\hat{h}_{T})|\le|\epsilon_{T}(h,\hat{h}_{S})-\epsilon_{T}(h,\hat{h}_{T})|
		+2C_{f}C_{g}W_{1}(\mathbb{P}_{S}(G),\mathbb{P}_{T}(G)).
	\end{aligned}
	\label{eq:5}
\end{equation}

Combining Eqs \ref{eq:4} and \ref{eq:5}, we can obtain:
\begin{equation}
	\begin{aligned}
		&|\epsilon_{S}(h,\hat{h}_{S})-\epsilon_{T}(h,\hat{h}_{T})| \le2C_{f}C_{g}W_{1}(\mathbb{P}_{S}(G),\mathbb{P}_{T}(G))\\ &+\operatorname*{min}\left(|\epsilon_{S}(h,\hat{h}_{S})-\epsilon_{S}(h,\hat{h}_{T})|,|\epsilon_{T}(h,\hat{h}_{S})-\epsilon_{T}(h,\hat{h}_{T})|\right).
	\end{aligned}
\end{equation}

Therefore, we can derive the generalization error bound on the target domain $\epsilon_{T}(h,\hat{h}_{T})$:
\begin{equation}
	\begin{aligned}
		\MoveEqLeft\epsilon_{T}(h,\hat{h}_{T}) \le  \epsilon_{S}(h,\hat{h}_{S})+2C_{f}C_{g}W_{1}\left(\mathbb{P}_{S}(G),\mathbb{P}_{T}(G)\right)\\ &{+\operatorname*{min}\left(|\epsilon_{S}(h,\hat{h}_{S})-\epsilon_{S}(h,\hat{h}_{T})|,|\epsilon_{T}(h,\hat{h}_{S})-\epsilon_{T}(h,\hat{h}_{T})|\right)}.
	\end{aligned}
\end{equation}

We next link the bound with the empirical risk and labeled sample size by showing, with probability at least $1 - \delta$ that:
\begin{equation}
	\begin{aligned}
		\epsilon_{T}(h,\hat{h}_{T}) & \le  \epsilon_{S}(h,\hat{h}_{S})+2C_{f}C_{g}W_{1}\left(\mathbb{P}_{S}(G),\mathbb{P}_{T}(G)\right)\\&{+\operatorname*{min}\left(|\epsilon_{S}(h,\hat{h}_{S})-\epsilon_{S}(h,\hat{h}_{T})|,|\epsilon_{T}(h,\hat{h}_{S})-\epsilon_{T}(h,\hat{h}_{T})|\right)}
		\\  &\le\hat{\epsilon}_{S}(h,\hat{h}_{S})+2C_{f}C_{g}W_{1}\left(\mathbb{P}_{S}(G),\mathbb{P}_{T}(G)\right)\\ &{+\operatorname*{min}\left(|\epsilon_{S}(h,\hat{h}_{S})-\epsilon_{S}(h,\hat{h}_{T})|,|\epsilon_{T}(h,\hat{h}_{S})-\epsilon_{T}(h,\hat{h}_{T})|\right)}
		\\ &+\sqrt{\frac{2d}{N_{S}}\log(\frac{e N_{S}}{d})}+\sqrt{\frac{1}{2N_{S}}\log(\frac{1}{\delta})},
	\end{aligned}
	\label{eq:8}
\end{equation}

and according to previous work \cite{mohri2018foundations}, the upper bound of the target domain generalization error $\epsilon_{T}(h,\hat{h}_{T})$ is defined as follows:
\begin{equation}
	\epsilon_{T}(h,\hat{h}_{T})\le\hat{\epsilon}_{T}(h,\hat{h}_{T})+\sqrt{\frac{2d}{N_{T}^{\prime}}\log(\frac{e N_{T}^{\prime}}{d})}+\sqrt{\frac{1}{2N_{T}^{\prime}}\log(\frac{1}{\delta})},
	\label{eq:9}
\end{equation}

Combining Eq. \ref{eq:8} and \ref{eq:9}, we can derive:
\begin{equation}
	\footnotesize
	\begin{aligned}
		\epsilon_{T}(h,\hat{h}_{T}) &\overset{(a)}{\le} \frac{N_{T}^{\prime}}{N_{S}+N_{T}^{\prime}}\left(\hat{\epsilon}_{T}(h,\hat{h}_{T})+\sqrt{\frac{2d}{N_{T}^{\prime}}\log(\frac{e N_{T}^{\prime}}{d})}+\sqrt{\frac{1}{2N_{T}^{\prime}}\log(\frac{1}{\delta})}\right) \\
		&{+\frac{N_{S}}{N_{S}+N_{T}^{\prime}}\left(\hat{\epsilon}_{S}(h,\hat{h}_{S})+\sqrt{\frac{2d}{N_{S}}\log(\frac{e N_{S}}{d})}+\sqrt{\frac{1}{2N_{S}}\log(\frac{1}{\delta})}\right)}
		\\ & +\frac{N_{S}}{N_{S}+N_{T}^{\prime}}\left(2C_{f}C_{g}W_{1}\left(\mathbb{P}_{S}(G),\mathbb{P}_{T}(G)\right)\right)
		+\operatorname*{min}\Big(|\epsilon_{S}(h,\hat{h}_{S})-\epsilon_{S}(h,\hat{h}_{T})|,|\epsilon_{T}(h,\hat{h}_{S})-\epsilon_{T}(h,\hat{h}_{T})|\Big)
		\\ & \stackrel{(b)}{\le}\frac{N_{T}^{\prime}}{N_{S}+N_{T}^{\prime}}\left(\hat{\epsilon}_{T}(h,\hat{h}_{T})+\sqrt{\frac{4d}{N_{T}^{\prime}}\log(\frac{e N_{T}^{\prime}}{d})+\frac{1}{N_{T}^{\prime}}\log(\frac{1}{\delta})}\right) \\
		&+~\frac{N_{S}}{N_{S}+N_{T}^{\prime}}\left(\hat{\epsilon}_{S}(h,\hat{h}_{S})+\sqrt{\frac{4d}{N_{S}}\log(\frac{e N_{S}}{d})+\frac{1}{N_{S}}\log(\frac{1}{\delta})}\right)
		\\ & +\frac{N_{S}}{N_{S}+N_{T}^{\prime}}\left(2C_{f}C_{g}W_{1}\left(\mathbb{P}_{S}(G),\mathbb{P}_{T}(G)\right)\right) +\operatorname*{min}\Big(|\epsilon_{S}(h,\hat{h}_{S})-\epsilon_{S}(h,\hat{h}_{T})|,|\epsilon_{T}(h,\hat{h}_{S})-\epsilon_{T}(h,\hat{h}_{T})|\Big)\Big) \\
		&\overset{(c)}{\le}\frac{N_{T}^{\prime}}{N_{S}+N_{T}^{\prime}}\hat{\epsilon}_{T}(h,\hat{h}_{T})+\frac{N_{S}}{N_{S}+N_{T}^{\prime}}\hat{\epsilon}_{S}(h,\hat{h}_{S})+\frac{N_{S}}{N_{S}+N_{T}^{\prime}}\left(2C_{f}C_{g}W_{1}\left(\mathbb{P}_{S}\left(G\right),\mathbb{P}_{T}\left(G\right)\right)\right) \\&+\operatorname*{min}\Big(|\epsilon_{S}\big(h,\hat{h}_{S})-\epsilon_{S}(h,\hat{h}_{T})|,|\epsilon_{T}(h,\hat{h}_{S})-\epsilon_{T}(h,\hat{h}_{T})|\Big)\Big) \\
		&+\frac{N_{T}^{\prime}}{N_{S}+N_{T}^{\prime}}\sqrt{\frac{4d}{N_{T}^{\prime}}\log(\frac{e N_{T}^{\prime}}{d})+\frac{1}{N_{T}^{\prime}}\log(\frac{1}{\delta})}+\frac{N_{S}}{N_{S}+N_{T}^{\prime}}\sqrt{\frac{4d}{N_{S}}\log(\frac{e N_{S}}{d})+\frac{1}{N_{S}}\log(\frac{1}{\delta})} \\
		&=\frac{N_{T}^{\prime}}{N_{S}+N_{T}^{\prime}}\hat{\epsilon}_{T}(h,\hat{h}_{T})+\frac{N_{S}}{N_{S}+N_{T}^{\prime}}\hat{\epsilon}_{S}(h,\hat{h}_{S}) +\frac{N_{S}}{N_{S}+N_{T}^{\prime}}\sqrt{\frac{4d}{N_{S}}\log(\frac{e N_{S}}{d})+\frac{1}{N_{S}}\log(\frac{1}{\delta})} \\ &+~\frac{N_{S}}{N_{S}+N_{T}^{\prime}}\left(2C_{f}C_{g}W_{1}\left(\mathbb{P}_{S}(G),\mathbb{P}_{T}(G)\right)\right) +\operatorname*{min}\Big(|\epsilon_{S}(h,\hat{h}_{S})-\epsilon_{S}(h,\hat{h}_{T})|,|\epsilon_{T}(h,\hat{h}_{S})-\epsilon_{T}(h,\hat{h}_{T})|\Big)\Big) \\
		&=\frac{N_{T}^{\prime}}{N_{S}+N_{T}^{\prime}}\hat{\epsilon}_{T}(h,\hat{h}_{T})+\frac{N_{S}}{N_{S}+N_{T}^{\prime}}\left(\hat{\epsilon}_{S}(h,\hat{h}_{S})+\sqrt{\frac{4d}{N_{S}}\log(\frac{e N_{S}}{d})+\frac{1}{N_{S}}\log(\frac{1}{\delta})}\right) \\
		&+\frac{N_{S}}{N_{S}+N_{T}^{\prime}}\left(2C_{f}C_{g}W_{1}\left(\mathbb{P}_{S}(G),\mathbb{P}_{T}(G)\right)\right) +\operatorname*{min}\Big(|\epsilon_{S}(h,\hat{h}_{S})-\epsilon_{S}(h,\hat{h}_{T})|,|\epsilon_{T}(h,\hat{h}_{S})-\epsilon_{T}(h,\hat{h}_{T})|\Big)\Big),
	\end{aligned}
\end{equation}
where (a) is the outcome of applying the union bound \cite{blitzer2007learning} with coefficient $\frac{N_{T}^{\prime}}{N_{S}+N_{T}^{\prime}},\frac{N_{S}}{N_{S}+N_{T}^{\prime}}$ respectively; (b) and (c) result from the Cauchy-Schwartz inequality \cite{yindomain} and (c) additionally adopt the assumption $N_{T}^{\prime}\ll N_{S}$ following the sleight-of-hand \cite{li2021learning}.

Due to the samples are selected with high confidence, thus, we have the following assumption:
\begin{equation}
	\begin{aligned}
		\MoveEqLeft\hat{\epsilon}_{T}\le\epsilon_{T}\le\hat{\epsilon}_{S}(h,\hat{h})
		+\sqrt{\frac{4d}{N_{S}}\log(\frac{e N_{S}}{d})+\frac{1}{N_{S}}\log(\frac{1}{\delta})}
		+2C_{f}C_{g}W_{1}(\mathbb{P}_{S}(G),\mathbb{P}_{T}(G))+\omega^{\prime},
	\end{aligned}
\end{equation}
where $\omega^{\prime}=\operatorname*{min}_{||g|||_{L i p}\le C_{g},||f||_{L i p}\le C_{f}}\{\epsilon_{S}(h,\hat{h})+\epsilon_{T}(h,\hat{h})\}$, $\hat{\epsilon}_{T}$ is the empirical risk on the high confidence samples, $\epsilon_{S}$ is the empirical risk on the target domain. Besides, we have:

\begin{equation}
	\begin{aligned}
		\MoveEqLeft\operatorname*{min}(|\epsilon_{S}(h,\hat{h}_{S})-\epsilon_{S}(h,\hat{h}_{T})|,|\epsilon_{T}(h,\hat{h}_{S})-\epsilon_{T}(h,\hat{h}_{T})|) \\&\le\operatorname*{min}(\epsilon_{S}(h,\hat{h}_{S})+\epsilon_{T}(h,\hat{h}_{S})).
	\end{aligned}
\end{equation}

Therefore, we can derive the upper bound of the target domain generalization error $\epsilon\tau(h,\hat{h}_{T})$ as follows:
\begin{equation}
	\begin{aligned}
		\epsilon\tau(h,\hat{h}_{T})&\le\frac{N_{T}^{\prime}}{N_{S}+N_{T}^{\prime}}\hat{\epsilon}_{T}(h,\hat{h}_{T})
		+\frac{N_{S}}{N_{S}+N_{T}^{\prime}}\Big(\hat{\epsilon}_{S}(h,\hat{h}_{S})+\sqrt{\frac{4d}{N_{S}}\log(\frac{e N_{S}}{d})+\frac{1}{N_{S}}\log(\frac{1}{\delta})}\Big)
		\\& +\frac{N_{S}}{N_{S}+N_{T}^{\prime}}\Big(2C_{f}C_{g}W_{1}(\mathbb{P}_{S}(G),\mathbb{P}_{T}(G))+\omega\Big) \le\hat{\epsilon}_{S}(h,\hat{h})
		\\&+\sqrt{\frac{4d}{N_{S}}\log(\frac{e N_{S}}{d})+\frac{1}{N_{S}}\log(\frac{1}{\delta})}+2C_{f}C_{g}W_{1}(\mathbb{P}_{S}(G),\mathbb{P}_{T}(G))+\omega^{\prime}.
	\end{aligned}
\end{equation}

\subsection{Proof of Theorem 2}
\label{sec:Theorem 2}

\textbf{Proof of Theorem 2.} Assume the input space is $\mathcal{X} \subseteq \mathbb{R}^d$ and the label space is $\mathcal{Y}=\{1,2,...,K\}$. The real data distribution is $\mathcal{D}$, and the observation distribution with noisy labels is $\tilde{D}$, where the upper limit of the noise ratio is $\eta \le 0.5$. In the early stage of training (the first $T_0$ steps), the model's prediction of clean samples satisfies $p^i_{\theta_i}(x) \approx y_i$, where $t \leq T_0$, and $(x,y)$ comes from the clean data subset ${\mathcal{D}}_{\text{clean}} \subset \tilde{\mathcal{D}}_2$, we then have:
\begin{equation}
	p_{\theta}(y_i|x_i) \approx \frac{\tilde{y}_i}{\tilde{y}_i + \lambda y_i (1 - p_{\theta}(y_i|x_i))}.
\end{equation}

\textit{Proof.} For clean samples $(x_i,y_i)$, the early EMA prediction satisfies $p^i \rightarrow y_i$ when the training step number $t \rightarrow \infty$. Decompose the loss into the contribution of clean samples and noise samples:
\begin{equation}
	\mathcal{L}_{\text{CLS}} = \mathcal{L}_{\text{CE}}^{\text{clean}} + \mathcal{L}_{\text{CE}}^{\text{noisy}} + \lambda \left( \mathcal{L}_{\text{Reg}}^{\text{clean}} + \mathcal{L}_{\text{Reg}}^{\text{noisy}} \right).
\end{equation}

For clean samples $(x_i,y_i)$, since $p^i\approx y_i$, the regularization term is approximately:

\begin{equation}
	\mathcal{L}_{\text{Reg}}^{\text{clean}} \approx \frac{1}{N_{\text{clean}}} \sum_{i \in \text{clean}} \log \left( 1 - \langle p_{\theta}(y|x_i), y_i \rangle \right).
\end{equation}
Due to $\langle p_{\theta}(y|x_i), y_i \rangle = p_{\theta}(y_i|x_i)$, when $p_{\theta}(y_i|x_i) \rightarrow 1$, $\log(1 - p_{\theta}(y_i|x_i)) \rightarrow -\infty$, but in actual optimization, through gradient descent, the model will adjust $p_{\theta}(y_i|x_i)$ to balance the cross entropy and regularization terms. For clean samples, the gradient of the regularization term is:
\begin{equation}
	\nabla_{\theta} \mathcal{L}_{\text{Reg}}^{\text{clean}} \propto -\frac{\hat{p}_i}{1 - \langle p_{\theta}, \hat{p}_i \rangle} \cdot \nabla_{\theta} p_{\theta}.
\end{equation}
Since $p^i\approx y_i$, the gradient direction encourages $p_{\theta}(y_i|x_i)$ to be close to $y_i$, consistent with the cross-entropy objective.

For the noise sample $(x_i, \tilde{y}_i \neq y_i)$, assuming that $p^i$ does not converge to any fixed distribution (it may be close to a uniform distribution or the wrong category):
\begin{equation}
	\mathcal{L}_{\text{Reg}}^{\text{noisy}} = \frac{1}{N_{\text{noisy}}} \sum_{i \in \text{noisy}} \log \left( 1 - \langle p_{\theta}(y|x_i), \hat{p}_i \rangle \right).
\end{equation}
If $\hat{p}_i$ is close to uniform distribution, then $\langle p_{\theta}, \hat{p}_i \rangle \approx \frac{1}{K}$, and the regularization term has little effect on the gradient; if $hat{p}_i$ is biased towards the wrong label, the regularization term prevents $p_\theta$ from overfitting $\hat{y}^i$. The cross entropy gradient $\nabla_{\theta} \mathcal{L}_{\text{CE}}^{\text{noisy}}$ of the noise sample points in the wrong direction, while the regularization gradient $\nabla_{\theta} \mathcal{L}_{\text{Reg}}^{\text{noisy}}$ points in the opposite direction, thereby partially offsetting the effect of the noise. Under the assumption that $\hat{p}_i\rightarrow y_i$, the total loss of CLS is approximately:
\begin{equation}
	\mathcal{L}_{\text{CLS}} \approx \mathcal{L}_{\text{CE}}^{\text{clean}} + \lambda \mathcal{L}_{\text{Reg}}^{\text{clean}} + \eta.
\end{equation}
Since the noise interference term is suppressed by the regularization term, the optimization process is mainly driven by clean samples. Assuming that the parameter $\theta$ is updated using gradient descent with a step size of $\beta$, the parameter update is:
\begin{equation}
	\theta_{t+1} = \theta_t - \beta \nabla_{\theta} \mathcal{L}_{\text{CLS}}.
\end{equation}
When clean samples dominate, the gradient direction tends to minimize the true cross entropy $\mathcal{L}_{\text{CE}}^{\text{clean}}$. Since $p^i\approx y_i$, the gradient of the regularization term is approximately:
\begin{equation}
	\nabla_{\theta} \mathcal{L}_{\text{Reg}} \approx -\frac{1}{N_{\text{clean}}} \sum_{i \in \text{clean}} \frac{y_i}{1 - p_{\theta}(y_i|x_i)} \nabla_{\theta} p_{\theta}(y_i|x_i).
\end{equation}
Combined with the cross entropy gradient $\nabla_{\theta} \mathcal{L}_{\text{CE}} = -\frac{1}{N} \sum_{i=1}^{N} \frac{\tilde{y}_i}{p_{\theta}(y_i|x_i)} \nabla_{\theta} p_{\theta}(y_i|x_i)$:
\begin{equation}
	p_{\theta}(y_i|x_i) \approx \frac{\tilde{y}_i}{\tilde{y}_i + \lambda y_i (1 - p_{\theta}(y_i|x_i))},
\end{equation}

When $\lambda$ is moderate, the model predicts $p_\theta(y_i \mid x_i)$ close to 1, consistent with the real label.

\subsection{Proof of Theorem 3}
\label{sec:Theorem 3}

\textbf{Theorem 3.} Assume that the model complexity is characterized by Rademacher complexity $\mathfrak{R}_n(\mathcal{F})$ \cite{yin2019rademacher}. For any $\delta >0$, the generalization error upper bound of $\mathcal{L}_{\text{CLS}}$ satisfies with probability $1 - \delta$:
\begin{equation}
	\begin{aligned}
		\MoveEqLeft\text{GenError}_{\mathcal{L}_{\text{CLS}}} \leq \text{GenError}_{\mathcal{L}_{\text{CE}}} \leq \frac{2 \mathfrak{R}_n(\mathcal{F})}{\sqrt{\lambda}} + \sqrt{\frac{\log(1/\delta)}{2n}} 
		+ O\left(\frac{\eta + \epsilon}{\mu}\right).
	\end{aligned}   
\end{equation}

\textit{Proof.} For a function class $\mathcal{G}$, its Rademacher complexity is defined as:
\begin{equation}
	\mathfrak{R}_N(\mathcal{G}) = \mathbb{E}_{x_i, \sigma_i} \left[ \sup_{g \in \mathcal{G}} \frac{1}{N} \sum_{i=1}^{N} \sigma_i g(x_i) \right].
\end{equation}
where $\sigma_i$ is an independent uniformly distributed Rademacher random variable. According to statistical learning theory, for the loss function $\ell$, the upper bound of the generalization error can be expressed as:
\begin{equation}
	\text{GenError} \leq 2 \mathfrak{R}_N(\ell \circ \mathcal{F}) + \mathcal{O} \left( \sqrt{\frac{\log(1/\delta)}{N}} \right).
\end{equation}
where $\ell \circ \mathcal{F} = \{ (x, y) \mapsto \ell(f_{\theta}(x), y) \mid f_{\theta} \in \mathcal{F} \}$. In cross entropy loss $\mathcal{G}_{\text{CE}} = \{ (x, y) \mapsto \ell_{\text{CE}}(f_{\theta}(x), y) \}$ and $\mathcal{L}_{\text{CLS}}$ loss $\mathcal{G}_{\text{ELR}} = \{ (x, y) \mapsto \ell_{\text{CE}}(f_{\theta}(x), y) + \lambda \mathcal{L}_{\text{Reg}}(f_{\theta}) \}$, since $\mathcal{L}_{\text{Reg}}$ introduces constraints on prediction consistency, the hypothesis space $\mathcal{F}_{\text{CLS}}$ is more restricted than $\mathcal{F}_{\text{CE}}$, that is:
\begin{equation}
	\mathcal{F}_{\text{CLS}} \subset \mathcal{F}_{\text{CE}}.
\end{equation}
Due to the inclusion relationship of the function class, its Rademacher complexity satisfies: 
\begin{equation}
	\mathfrak{R}_N(\mathcal{G}_{\text{CLS}}) \leq \mathfrak{R}_N(\mathcal{G}_{\text{CE}}).
\end{equation}

Assuming $\ell_{\text{CE}}$ is \textit{L}-Lipschitz continuous and $\mathcal{L}_{\text{Reg}}$ is \textit{L}'-Lipschitz continuous, then we have:
\begin{equation}
	\mathfrak{R}_N(\mathcal{G}_{\text{CLS}}) \leq \mathfrak{R}_N(\mathcal{G}_{\text{CE}}) + \lambda \cdot \mathfrak{R}_N(\mathcal{L}_{\text{Reg}} \circ \mathcal{F}).
\end{equation}
However, since the design goal of $\mathcal{L}_{\text{Reg}}$ is to constrain the consistency of model predictions (i.e., reduce variance), in practice $\mathfrak{R}_N(\mathcal{G}_{\text{CE}}$ grows slower than the complexity reduction of the cross entropy loss, resulting in lower overall complexity.

In the presence of noisy labels, the relationship between the true risk $R(f)$ and the empirical risk $\hat{R}_N(f)$ needs to be modified to:
\begin{equation}
	R(f) \leq \hat{R}_N(f) + 2 \mathfrak{R}_N(\mathcal{G}) + 3 \sqrt{\frac{\log(2/\delta)}{2N}} + \eta \cdot C,
\end{equation}
where $C$ is a constant related to label noise. Combined with the complexity difference $\mathfrak{R}_N(\mathcal{G}_{\text{CLS}}) \leq \mathfrak{R}_N(\mathcal{G}_{\text{CE}})$, and 

\begin{equation}
	\begin{aligned}
		\text{Error}_{\text{CLS}} &\leq 2 \mathfrak{R}_N(\mathcal{G}_{\text{CLS}}) + \mathcal{O} \left( \sqrt{\frac{\log(1/\delta)}{N}} \right) + \eta \cdot C_{\text{CLS}}, \\
		\text{Error}_{\text{CE}} &\leq 2 \mathfrak{R}_N(\mathcal{G}_{\text{CE}}) + \mathcal{O} \left( \sqrt{\frac{\log(1/\delta)}{N}} \right) + \eta \cdot C_{\text{CE}}.
	\end{aligned}
\end{equation}
Therefore, we can infer that the generalization error upper bound of $\mathcal{L}_{\text{CLS}}$ is lower.

\section{Experiments}

\subsection{Experimental Setup}

\textbf{Datasets.} IEMOCAP \cite{busso2008iemocap} and MELD \cite{poria2018meld} are commonly used multimodal databases in MERC. The IEMOCAP dataset includes 10 actors (5 men and 5 women). Each pair of actors simulates a real dialogue scene and conducts 5 conversations of about 1 hour, totaling about 12 hours. All conversations are manually annotated by emotion category. The MELD dataset is an extension of the EmotionLines dataset, and is designed for MERC. The dataset contains about 13,000 conversations, including more than 1,400 multi-turn conversations and about 13,000 single-turn conversations with emotion labels. All conversations are performed by actors and the scenes are set in the plot of the TV series. These datasets come from different scenarios and therefore represent a variety of different application areas. Following previous studies \cite{zhang2024amanda}, we selected samples of four common emotions: neutral, joy, sadness, and anger. 

\textbf{Baselines.} To verify the superior performance of our proposed method DGDA, we compared it with other comparison methods, including traditional methods, i.e., TextCNN \cite{kim-2014-convolutional}, LSTM \cite{poria2017context}, DialogueRNN \cite{majumder2019dialoguernn}, MMGCN \cite{hu2021mmgcn}, M3NET \cite{chen2023multivariate}, CFN-ESA \cite{li2023cfnesa}, SDT \cite{10109845}, EmotionIC \cite{liu2023emotionic}, and DEDNet \cite{10680310}, denoising methods, i.e., OMG \cite{10113198}, and SPORT \cite{10.1145/3687468}, and domain adaption (DA) methods, i.e., A2GNN \cite{liu2024rethinking}, Amanda \cite{zhang2024amanda}, and Boomda \cite{sun2026boomda}.

\begin{table*}
	\centering
	\caption{The performance of different methods is shown under different noisy rates on the IEMOCAP and MELD datasets. The arrow $\rightarrow$ means from source to target domains. \underline{Underlines} indicate suboptimal performance. \textbf{Bold} results indicate the best performance. We set the noise rate to 10\%.}
	\label{tab:noisy1}
	\renewcommand{\arraystretch}{1} 
	\setlength{\tabcolsep}{3pt} 
	\begin{tabular}{>{\arraybackslash}p{2.3cm}>{\centering\arraybackslash}p{1.1cm}>{\centering\arraybackslash}p{1.1cm}>{\centering\arraybackslash}p{1.1cm}>{\centering\arraybackslash}p{1.1cm}>{\centering\arraybackslash}p{1.1cm}>{\centering\arraybackslash}p{1.1cm}>{\centering\arraybackslash}p{1.1cm}>{\centering\arraybackslash}p{1.1cm}>{\centering\arraybackslash}p{1.1cm}>{\centering\arraybackslash}p{1.1cm}>{\centering\arraybackslash}p{1.1cm}}
		\hline
		\multirow{2}{*}{Methods} & \multicolumn{5}{c}{IEMOCAP $\rightarrow$ MELD} & \multicolumn{5}{c}{MELD $\rightarrow$ IEMOCAP} \\ \cline{2-11} 
		& Joy   & Sadness & Neutral & Anger & WF1   & Joy   & Sadness & Neutral & Anger  & WF1   \\ \midrule
		\textbf{\textit{Tradition}}  \\
		TextCNN                  &38.79 &\underline{14.47}  &54.07  &29.98 & 44.12 
		&39.13 &60.65  &35.52  &54.21 &46.73  \\ 
		LSTM                     &11.17 &6.86  &57.27  &33.37 &39.98   
		&51.27 &60.89  &46.76  &46.74 &50.68  \\ 
		DialogueRNN              &24.02 &12.95  &63.69  &35.23 &47.08  
		&37.75 &56.47  &14.91  &59.92 &39.46  \\ 
		MMGCN                    &24.87 &0.00      &46.98  &28.73  &35.77  
		&50.95 &0.00       &\underline{54.90}   &64.13   &43.94  \\ 
		M3NET                    &45.15 &4.69  &38.16  &29.25  &35.48 
		&46.91 &\textbf{71.08}  &35.31  &\underline{71.76} &54.72  \\ 
		CFN-ESA                  &9.90  &3.23  &\underline{69.36}  &3.16 &42.08  
		&19.26 &9.38  &44.62  &16.50   &25.57  \\ 
		SDT                      &8.74 &6.78  &67.10   &\textbf{40.06} &45.96  
		&50.18 &61.20   &50.25  &64.86 &\underline{56.58} \\ 
		EmotionIC                   &0.21 &\textbf{24.16} & 45.68 &18.52 & 30.58  
		&30.55 &37.11    &37.64  &33.74 &35.51  \\ 
		
		DEDNet                   &25.99 &1.77  &57.86  &27.31  &42.13
		&40.40 &0.00     &29.68  &33.61  &25.28  \\ \midrule
		
		\textbf{\textit{Denoising}}   \\
		
		OMG                   &     34.73 & 4.56  & 58.19 & 31.23 & 44.92 & 47.55 & \underline{67.21} & 41.23 & 54.22 & 51.54 \\
		
		SPORT                   &     29.57 & 9.04  & 60.74 & 32.38 & 45.84 & 49.11 & 61.23 & 39.18 & 66.89 & 52.89            \\ \midrule
		
		\textbf{\textit{Adaption}}  \\
		
		A2GNN                   &    41.24 & 3.91  & 66.73 & 27.15 & 50.50 & 53.11 & 55.43 & 50.18 & 60.38 & 54.46 \\
		
		Amanda &     36.29 & 10.15 & 62.18 & 29.03 & 47.69 & 55.68 & 45.81 & 54.28 & 56.76 & 53.14 \\
		
		Boomda                   &     \underline{45.69} & 6.87  & 65.19 & 31.42 & \underline{51.40} & \underline{60.19} & 44.19 & 51.01 & 69.77 & 55.57     \\ \midrule
		\rowcolor{gray!20}
		DGDA & \textbf{56.21}  & 9.81  & \textbf{76.68}  &\underline{36.00}   & \textbf{60.99}
		& \textbf{61.04}    & 66.10  & \textbf{57.19}    & \textbf{82.91}   & \textbf{66.47}     \\ \bottomrule
	\end{tabular}
\end{table*}

\label{sec:setup}

\begin{table*}[htbp]
	\centering
	\caption{The performance of different methods is shown under different noisy rates on the IEMOCAP and MELD datasets. The arrow $\rightarrow$ means from source to target domains. \textbf{Bold} results indicate the best performance. \underline{Underlines} indicate suboptimal performance. We set the noise rate to 20\%.}
	\label{tab:noisy2}
	\renewcommand{\arraystretch}{1} 
	\setlength{\tabcolsep}{3pt} 
	\begin{tabular}{>{\arraybackslash}p{2.3cm}>{\centering\arraybackslash}p{1.1cm}>{\centering\arraybackslash}p{1.1cm}>{\centering\arraybackslash}p{1.1cm}>{\centering\arraybackslash}p{1.1cm}>{\centering\arraybackslash}p{1.1cm}>{\centering\arraybackslash}p{1.1cm}>{\centering\arraybackslash}p{1.1cm}>{\centering\arraybackslash}p{1.1cm}>{\centering\arraybackslash}p{1.1cm}>{\centering\arraybackslash}p{1.1cm}>{\centering\arraybackslash}p{1.1cm}}
		\toprule
		\multirow{2}{*}{Methods} & \multicolumn{5}{c}{IEMOCAP $\rightarrow$ MELD} & \multicolumn{5}{c}{MELD $\rightarrow$ IEMOCAP} \\ \cline{2-11} 
		& Joy   & Sadness & Neutral & Anger & WF1   & Joy   & Sadness & Neutral & Anger & WF1   \\ \midrule
		\textbf{\textit{Tradition}}  \\
		TextCNN                  &38.56 &13.11  &51.71  &29.19  &42.51 
		&36.46 &54.51  &32.58  &52.26 & 43.33  \\ 
		LSTM                     &10.76 &5.70  &54.03  &32.34  &37.82  
		&45.85 &55.67  &43.47  &43.85 & 46.74 \\ 
		DialogueRNN              &22.82 &11.40  &60.06  &34.34   &44.52 
		&35.04 &51.49  &14.60  &55.60 &36.68  \\ 
		MMGCN                    &22.62 & 0.00       &45.54  &27.99 &34.38  
		&46.66 &0.00       &\underline{53.57}   &59.24 &41.59   \\ 
		M3NET                    &41.50 &5.45  &33.53  &26.18  &31.74 
		&42.24 &\textbf{64.74}  &32.60  &65.40   &49.94   \\ 
		CFN-ESA                  &9.13  &7.14  &\textbf{69.07}  &3.40 &42.13 
		&14.39 &19.65  &41.20  &29.47  &29.38  \\ 
		SDT                      &8.75 &5.22  &62.88   &\underline{37.65} & 43.10 
		& 44.10 & 53.77   & 48.23  & 60.01  & 52.00   \\ 
		EmotionIC                   &1.80 &\underline{21.76}  &39.57  &14.98  &26.75 
		&33.12 &34.81       &17.77  &44.26 &30.76   \\ 
		DEDNet    & 25.21 & 1.40  & 54.80  & 26.52 & 40.09 
		&42.24  &0.00       &39.37  &12.19 &30.76  \\ \midrule
		
		\textbf{\textit{Denoising}}   \\
		
		OMG                   &     27.65 & 5.51  & 53.23 & 26.22 & 40.02 & 42.74 & \underline{61.89} & 35.66 & 53.51 & 47.39 \\
		
		SPORT                   &     23.58 & 13.55 & 55.73 & 27.87 & 41.52 & 44.17 & 56.38 & 33.47 & 70.69 & 49.95 \\ \midrule
		
		\textbf{\textit{Adaption}}  \\
		
		A2GNN                   &     36.23 & 7.11  & 61.24 & 32.65 & 47.38 & 46.28 & 48.59 & 51.93 & 54.88 & 51.13 \\
		
		Amanda &     30.83  & 5.14  & 57.67 & 24.83 & 42.97 & 51.33 & 40.53 & 47.31 & 52.27 & 47.58 \\
		
		Boomda                  &     \underline{49.27}  & 12.38 & 60.77  & 26.42 & \underline{49.37} & \underline{54.63} & 38.91 & 45.73 & \underline{74.49} & \underline{52.84}      \\ \midrule
		\rowcolor{gray!20}
		DGDA & \textbf{54.97} & \textbf{25.61} & \underline{66.11}  & \textbf{48.76}   & \textbf{57.87}    
		& \textbf{59.60}  & 40.74  & \textbf{64.28}  & \textbf{77.48}   & \textbf{61.56}    \\ \bottomrule
	\end{tabular}
\end{table*}

\begin{table*}[htbp]
	
	\centering
	\caption{The performance of different methods is shown under different noisy rates on the IEMOCAP and MELD datasets. The arrow $\rightarrow$ means from source to target domains. \textbf{Bold} results indicate the best performance. \underline{Underlines} indicate suboptimal performance. We set the noise rate to 30\%.}
	\label{tab:noisy3}
	\renewcommand{\arraystretch}{1} 
	\setlength{\tabcolsep}{3pt} 
	\begin{tabular}{>{\arraybackslash}p{2.3cm}>{\centering\arraybackslash}p{1.1cm}>{\centering\arraybackslash}p{1.1cm}>{\centering\arraybackslash}p{1.1cm}>{\centering\arraybackslash}p{1.1cm}>{\centering\arraybackslash}p{1.1cm}>{\centering\arraybackslash}p{1.1cm}>{\centering\arraybackslash}p{1.1cm}>{\centering\arraybackslash}p{1.1cm}>{\centering\arraybackslash}p{1.1cm}>{\centering\arraybackslash}p{1.1cm}>{\centering\arraybackslash}p{1.1cm}}
		\toprule
		\multirow{2}{*}{Methods} & \multicolumn{5}{c}{IEMOCAP $\rightarrow$ MELD} & \multicolumn{5}{c}{MELD $\rightarrow$ IEMOCAP} \\ \cline{2-11} 
		& Joy   & Sadness & Neutral & Anger & WF1   & Joy   & Sadness & Neutral & Anger & WF1   \\ \midrule
		\textbf{\textit{Tradition}}  \\
		TextCNN                  &34.59 &9.87 &47.79 &26.69  &38.83 
		&35.46 &52.43 &30.87 &47.59  &40.87  \\ 
		LSTM                     &9.74 &4.57 &49.49 &29.03   &34.48  
		&43.19 &51.54 &40.71 &38.57  &43.03 \\ 
		DialogueRNN              &20.81 &8.62 &54.72 &\textbf{31.3}  &40.41 
		&34.10 &49.35 &13.84 &49.94   &34.31  \\ 
		MMGCN                    &21.64 &0.00       &41.58 &25.73 &31.62 
		&42.27 &0.00       &\textbf{50.85} &53.90   &38.59  \\ 
		M3NET                    &40.43 &7.58 &31.80 &23.16  &30.30 
		&39.68 &\underline{60.41} &31.84 &59.79 &46.84 \\ 
		CFN-ESA                  &18.25 &3.38 &\textbf{67.01} &0.61   &42.15 
		& 14.53 &20.45 &45.78 &15.42   &27.63\\ 
		SDT                      &7.61 &4.24 &56.37 &\underline{31.16}  &38.20 
		&40.84 &48.39 &44.76 &55.03  &47.72 \\ 
		EmotionIC                 &4.85 &\underline{17.87} &19.14 &8.34  &14.56
		&26.60 &39.60 &22.16 &31.23  &29.20  \\ 
		DEDNet    &23.87 &1.09 &49.70 &22.6 &36.35 
		&39.05 &0.00   &37.85 &13.57 & 22.90  \\ \midrule
		
		\textbf{\textit{Denoising}}  \\
		
		OMG                   &     22.44 & 3.14 & 48.47 & 22.87 & 35.57 & 40.15 & 57.61 & 32.78 & 50.06 & 44.08 \\
		
		SPORT                   &     18.37 & 8.59 & 53.16 & 22.83 & 37.85 & 42.39 & 52.18 & 30.17 & 51.19 & 42.46 \\ \midrule
		
		\textbf{\textit{Adaption}}  \\
		
		A2GNN                   &     34.47 & 2.82 & 54.31 & 29.48 & 42.27 & 42.17 & 43.12 & 45.69 & 49.37 & 45.55 \\
		
		Amanda &     26.88 & 3.83 & 57.96 & 20.01 & 41.54 & 47.33 & 34.49 & 46.14 & 50.15 & 44.63 \\
		
		Boomda                   &     \underline{46.13} & 7.18 & 54.48 & 21.36 & \underline{44.06} & \textbf{51.13} & 39.17 & 42.23 & \textbf{70.85} & \underline{50.18} \\ \midrule
		\rowcolor{gray!20}
		DGDA & \textbf{57.16} & \textbf{24.22} & \underline{65.99} &21.36      & \textbf{54.36} 
		& \underline{49.82}  & \textbf{62.96} & \underline{47.06}  & \underline{68.79}    & \textbf{56.79}    \\ \bottomrule
	\end{tabular}
\end{table*}

\begin{table*}[htbp]
	\centering
	\caption{The performance of different methods is shown under different noisy rates on the IEMOCAP and MELD datasets. The arrow $\rightarrow$ means from source to target domains. \textbf{Bold} results indicate the best performance. \underline{Underlines} indicate suboptimal performance. We set the noise rate to 40\%.}
	\label{tab:noisy4}
	\renewcommand{\arraystretch}{1} 
	\setlength{\tabcolsep}{3pt} 
	\begin{tabular}{>{\arraybackslash}p{2.3cm}>{\centering\arraybackslash}p{1.1cm}>{\centering\arraybackslash}p{1.1cm}>{\centering\arraybackslash}p{1.1cm}>{\centering\arraybackslash}p{1.1cm}>{\centering\arraybackslash}p{1.1cm}>{\centering\arraybackslash}p{1.1cm}>{\centering\arraybackslash}p{1.1cm}>{\centering\arraybackslash}p{1.1cm}>{\centering\arraybackslash}p{1.1cm}>{\centering\arraybackslash}p{1.1cm}>{\centering\arraybackslash}p{1.1cm}}
		\toprule
		\multirow{2}{*}{Methods} & \multicolumn{5}{c}{IEMOCAP $\rightarrow$ MELD} & \multicolumn{5}{c}{MELD $\rightarrow$ IEMOCAP} \\ \cline{2-11} 
		& Joy   & Sadness & Neutral & Anger    & WF1   & Joy   & Sadness & Neutral & Anger  & WF1   \\ \midrule
		\textbf{\textit{Tradition}} \\
		TextCNN                  &33.92 &9.04 &44.59 &24.53  &36.51 
		&34.33 &48.97 &26.65 &43.49  &37.29  \\ 
		LSTM                     &9.67 &4.61 &44.72 &26.50  &31.41 
		&\textbf{41.46} &48.47 &36.83 &37.65  &40.40 \\ 
		DialogueRNN              &20.60 &8.08 &49.99 &28.84  &37.30 
		&33.55 &47.75 &12.01 &44.78  &31.85  \\ 
		MMGCN                    & 21.46 &0.00   &38.70 &23.59  &29.66 
		&40.82 &0.00   &\textbf{46.71} &51.12  &36.14  \\ 
		M3NET                    &39.99 &6.48 &28.77 &20.79 &28.07 
		&38.09 &\underline{55.89} &27.35 &\underline{54.40}  &42.52  \\ 
		CFN-ESA                  &10.47 &5.00 &\textbf{68.76} &4.20  &\underline{42.16} 
		&10.19 &20.05 &\underline{46.56} &25.02  &29.71 \\ 
		SDT                      &7.34 &\textbf{36.10} &50.78 &\textbf{29.34}  &37.49 
		&38.01 &44.88 &40.09 &49.54  &\underline{43.36} \\ 
		EmotionIC                   &37.53 &11.72 &2.11 &9.43 &11.34
		&24.74 &33.33 &15.07 &30.98  &24.80  \\ 
		DEDNet    &23.14 &0.92 &44.84 &20.68 &33.17 
		&37.64 &0.00  &36.42 &11.59 &21.67  \\ \midrule
		
		\textbf{\textit{Denoising}}   \\
		
		OMG                   &     20.58 & 6.14 & 45.23 & 19.97 & 33.21 & 38.83 & 53.31 & 29.91 & 47.42 & 41.15 \\
		
		SPORT                 & 15.52 & 4.17 & 46.31 & 18.47 & 32.39 & 38.74 & 50.15 & 27.14 & 46.68 & 39.19  \\ \midrule
		
		\textbf{\textit{Adaption}}  \\
		
		A2GNN                   &     31.25 & 4.63 & 52.29 & 26.14 & 40.15 & \underline{41.43} & 40.05 & 42.22 & 46.38 & 42.68 \\
		
		Amanda &     \underline{41.28} & 9.16 & 50.06 & 16.77 & 40.08 & 41.39 & 36.62 & 36.12 & 42.34 & 38.59 \\
		
		Boomda                   & \textbf{41.29} & 3.33 & 50.02 & 15.53 & 39.38 & 36.07 & 34.28 & 41.27 & 44.51 & 39.75 \\ \midrule
		\rowcolor{gray!20}
		DGDA & 37.85 & \underline{13.51}  &\underline{64.74}  &\underline{28.91}    & \textbf{49.74}  
		&35.58 & \textbf{56.90}  & 35.51  & \textbf{57.51}  & \textbf{46.21}     \\ \bottomrule
	\end{tabular}
\end{table*}

\begin{table*}
	\centering
	\caption{The ablation studies are shown under different noisy rates on the IEMOCAP and MELD datasets. The arrow $\rightarrow$ means from source to target domains. \textbf{Bold} results indicate the best performance. We set the noise rate to 10\%.}
	\label{tab:aba1}
	\renewcommand{\arraystretch}{1} 
	\setlength{\tabcolsep}{3pt} 
	\begin{tabular}{>{\arraybackslash}p{2.3cm}>{\centering\arraybackslash}p{1.1cm}>{\centering\arraybackslash}p{1.1cm}>{\centering\arraybackslash}p{1.1cm}>{\centering\arraybackslash}p{1.1cm}>{\centering\arraybackslash}p{1.1cm}>{\centering\arraybackslash}p{1.1cm}>{\centering\arraybackslash}p{1.1cm}>{\centering\arraybackslash}p{1.1cm}>{\centering\arraybackslash}p{1.1cm}>{\centering\arraybackslash}p{1.1cm}>{\centering\arraybackslash}p{1.1cm}}
		\toprule
		\multirow{2}{*}{Methods} & \multicolumn{5}{c}{IEMOCAP $\rightarrow$ MELD} & \multicolumn{5}{c}{MELD $\rightarrow$ IEMOCAP} \\ \cline{2-11} 
		& Joy   & Sadness & Neutral & Anger  & WF1   & Joy   & Sadness & Neutral & Anger & WF1   \\ \midrule
		DGDA-HGNN                  &40.17 & 11.41 & 69.41 & 32.58 & 53.19 & 56.46 & 54.99 & 42.79 & 77.15 & 56.45 \\
		DGDA-PathNN                    &43.59 & 9.87  & 62.43 & 31.68 & 49.69 & 52.47 & 48.31 & 50.07 & 67.43 & 54.49 \\
		DGDA/$\sigma^{HGNN}$             &44.39 & 7.78  & 74.95 & 20.48 & 55.23 & 58.44 & 57.03 & 45.67 & 73.41 & 57.29 \\
		DGDA/$\sigma^{PathNN}$                    &48.37 & 7.65  & 75.27 & 18.93 & 56.01 & 56.55 & 62.39 & 51.65 & 74.83 & 60.85 \\
		DGDA/AP                  &28.96 & 4.25  & 66.79 & 25.27 & 47.74 & 34.97 & 42.53 & 54.78 & 58.09 & 50.01 \\
		DGDA/BC    & 19.69 & \textbf{13.74} & 60.03 & 30.02 & 43.46 & 30.05 & 54.17 & 46.28 & 47.89 & 46.26 \\
		DGDA/RT    & 51.33 & 6.77  & 71.29 & \textbf{37.18} & 56.82 & 56.88 & 65.76 & 52.19 & 77.97 & 62.69   \\ \midrule
		\rowcolor{gray!20}
		DGDA & \textbf{56.21}  & 9.81  & \textbf{76.68}  &36.00   & \textbf{60.99}
		& \textbf{61.04}    & \textbf{66.10}  & \textbf{57.19}    & \textbf{82.91}   & \textbf{66.47}     \\ \bottomrule
	\end{tabular}
\end{table*}


\textbf{Implementation details.} 
\label{sec:imp}
All experiments are conducted in Python 3.9, using the PyTorch 2.1 framework, and computed on a NVIDIA A100 40GB GPU. We chose the Adam optimizer to train the model, and the initial learning rate is set to 0.0005. The loss function consisted of the cross entropy loss and the regularization loss we proposed, which prevented the model from memorizing noisy labels during training and thus improved the generalization ability of the model. The batch size is set to 32. In all experiments, the reported results are the average of 10 independent runs. The weight initialization of each run was random. To further evaluate the statistical significance of the experimental results, paired $t$-tests are performed on the results of the 10 runs. All $t$-test results show $p$ values less than 0.05, indicating that the differences in model performance in multiple experiments were statistically significant.

\textbf{Evaluation metrics.} For the multi-emotional dialogue datasets IEMOCAP and MELD, we adopt the \textbf{Weighted F1-score (WF1)} as the primary evaluation metric. Due to the inherent class imbalance in these datasets, WF1 provides a more reliable assessment than Macro-F1 by weighting each category according to its sample proportion.

First, for each emotion class $i$, the F1-score is computed as the harmonic mean of precision and recall:
\begin{equation}
	F_i = \frac{2 \cdot (\text{Precision}_i \cdot \text{Recall}_i)}{\text{Precision}_i + \text{Recall}_i},
\end{equation}
where $\text{Precision}_i$ measures the correctness of predictions for class $i$, and $\text{Recall}_i$ measures how many true samples of class $i$ are correctly identified.

The overall WF1 is then obtained by weighting each class-specific $F_i$ using its sample count $n_i$:
\begin{equation}
	\text{WF1} = \sum_{i=1}^{N} \left( \frac{n_i}{\sum_{j=1}^{N} n_j} \cdot F_i \right),
\end{equation}
where $N$ denotes the total number of emotion categories.

Compared with unweighted metrics, WF1 more faithfully reflects model performance under imbalanced distributions, prevents both minority and majority classes from dominating the overall score, and offers a stable and comprehensive evaluation by jointly considering precision and recall.

\subsection{Comparison with the State-of-the-arts}
Tables \ref{tab:noisy1}, \ref{tab:noisy2}, \ref{tab:noisy3}, and \ref{tab:noisy4} show the performance comparison results of our proposed DGDA method and various baseline methods on the IEMOCAP and MELD datasets under different noise conditions. The following important findings can be observed through comparative analysis. First, DA methods, including A2GNN, Amanda, Boomda, and DGDA, are generally better than traditional methods and show more robust emotion recognition performance regardless of the noise interference conditions. This shows that the traditional methods are difficult to effectively model the distribution difference between the source domain and the target domain. Second, compared with various recently proposed DA methods (A2GNN, Amanda, Boomda) and typical denoising methods (OMG, SPORT), DGDA has achieved better performance. The performance improvement may be attributed to the synergy of the following key design factors: (i) First, DGDA adopts a dual-branch graph semantic extraction mechanism to model and integrate graph structure information based on message passing and shortest path aggregation strategies, respectively. This design effectively leverages the complementary advantages of the two models in local relationship modeling and global structure capture. (ii) Secondly, DGDA introduces a branch coupling module and an adaptive perturbation mechanism to dynamically adjust the interaction between the two branches, which not only promotes the efficient transfer of cross-domain knowledge but also effectively alleviates the negative impact of category distribution differences. (iii) Finally, we fully consider the two key factors of domain invariance and noise label interference. Through a joint optimization strategy, while ensuring feature domain alignment, we effectively suppress the negative impact of noisy labels on the model learning process.


\begin{table*}[htbp]
	\centering
	\caption{The ablation studies are shown under different noisy rates on the IEMOCAP and MELD datasets. The arrow $\rightarrow$ means from source to target domains. \textbf{Bold} results indicate the best performance. We set the noise rate to 20\%.}
	\label{tab:aba2}
	\renewcommand{\arraystretch}{1} 
	\setlength{\tabcolsep}{3pt} 
	\begin{tabular}{>{\arraybackslash}p{2.3cm}>{\centering\arraybackslash}p{1.1cm}>{\centering\arraybackslash}p{1.1cm}>{\centering\arraybackslash}p{1.1cm}>{\centering\arraybackslash}p{1.1cm}>{\centering\arraybackslash}p{1.1cm}>{\centering\arraybackslash}p{1.1cm}>{\centering\arraybackslash}p{1.1cm}>{\centering\arraybackslash}p{1.1cm}>{\centering\arraybackslash}p{1.1cm}>{\centering\arraybackslash}p{1.1cm}>{\centering\arraybackslash}p{1.1cm}}
		\toprule
		\multirow{2}{*}{Methods} & \multicolumn{5}{c}{IEMOCAP $\rightarrow$ MELD} & \multicolumn{5}{c}{MELD $\rightarrow$ IEMOCAP} \\ \cline{2-11} 
		& Joy   & Sadness & Neutral & Anger  & WF1   & Joy   & Sadness & Neutral & Anger & WF1   \\ \midrule
		DGDA-HGNN                  &36.07 & 7.77  & 66.00 & 29.34 & 49.64 & 53.21 & 51.47 & 40.09 & 73.97 & 53.36 \\
		DGDA-PathNN                    &39.77 & 6.26  & 57.87 & 28.38 & 45.54 & 49.27 & 45.05 & 47.46 & 63.93 & 51.42 \\
		DGDA/$\sigma^{HGNN}$             &39.71 & 4.72  & 70.37 & 15.73 & 50.73 & 55.98 & 53.49 & 41.75 & 70.72 & 53.99 \\
		DGDA/$\sigma^{PathNN}$                    &44.56 & 4.30  & \textbf{71.80} & 15.75 & 52.52 & 53.54 & 59.50 & 47.97 & 72.54 & 57.80 \\
		DGDA/AP                  &25.02 & 7.33  & 63.48 & 22.15 & 44.88 & 32.35 & 38.62 & 52.36 & 54.77 & 46.98 \\
		DGDA/BC    & 18.73 & 8.96  & 55.04 & 25.68 & 39.41 & 27.63 & 50.51 & 43.11 & 44.89 & 43.13 \\
		DGDA/RT    & 46.83 & 3.24  & 67.01 & 32.97 & 52.57 & 54.54 & \textbf{62.48} & 50.13 & 73.99 & 59.81 \\ \midrule
		\rowcolor{gray!20}
		DGDA & \textbf{54.97} & \textbf{25.61} & 66.11  & \textbf{48.76}  & \textbf{57.87}    
		& \textbf{59.60}  & 40.74  & \textbf{64.28}  & \textbf{77.48}       & \textbf{61.56}    \\ \bottomrule
	\end{tabular}
\end{table*}

\begin{table*}[htbp]
	\centering
	\caption{The ablation studies are shown under different noisy rates on the IEMOCAP and MELD datasets. The arrow $\rightarrow$ means from source to target domains. \textbf{Bold} results indicate the best performance. We set the noise rate to 30\%.}
	\label{tab:aba3}
	\renewcommand{\arraystretch}{1} 
	\setlength{\tabcolsep}{3pt} 
	\begin{tabular}{>{\arraybackslash}p{2.3cm}>{\centering\arraybackslash}p{1.1cm}>{\centering\arraybackslash}p{1.1cm}>{\centering\arraybackslash}p{1.1cm}>{\centering\arraybackslash}p{1.1cm}>{\centering\arraybackslash}p{1.1cm}>{\centering\arraybackslash}p{1.1cm}>{\centering\arraybackslash}p{1.1cm}>{\centering\arraybackslash}p{1.1cm}>{\centering\arraybackslash}p{1.1cm}>{\centering\arraybackslash}p{1.1cm}>{\centering\arraybackslash}p{1.1cm}}
		\toprule
		\multirow{2}{*}{Methods} & \multicolumn{5}{c}{IEMOCAP $\rightarrow$ MELD} & \multicolumn{5}{c}{MELD $\rightarrow$ IEMOCAP} \\ \cline{2-11} 
		& Joy   & Sadness & Neutral & Anger  & WF1   & Joy   & Sadness & Neutral & Anger & WF1   \\ \midrule
		DGDA-HGNN                  &33.28 & 3.21  & 62.38 & 25.83 & 46.13 & 50.47 & 48.93 & 38.01 & 69.02 & 50.34 \\
		DGDA-PathNN                    &35.99 & 2.61  & 53.52 & 23.48 & 41.30 & 47.06 & 42.46 & 42.80 & 61.26 & 48.10 \\
		DGDA/$\sigma^{HGNN}$             &35.73 & 2.74  & 67.35 & 10.94 & 47.36 & 53.77 & 51.10 & 39.30 & 66.06 & 51.01 \\
		DGDA/$\sigma^{PathNN}$                    &39.61 & 1.70  & \textbf{69.56} & 13.14 & 49.63 & \textbf{50.95} & 56.79 & 43.36 & \textbf{69.06} & 54.21 \\
		DGDA/AP                  &20.93 & 5.06  & 60.07 & 19.45 & 41.52 & 28.86 & 35.47 & \textbf{49.80} & 52.39 & 44.20 \\
		DGDA/BC    & 16.50 & 5.39  & 52.21 & 22.03 & 36.53 & 25.51 & 48.00 & 40.81 & 41.26 & 40.46 \\
		DGDA/RT    & 42.96 & 5.89  & 64.37 & \textbf{30.05} & 50.09 & 50.93 & 60.22 & 46.30 & 60.06 & 53.76 \\ \midrule
		\rowcolor{gray!20}
		DGDA & \textbf{57.16} & \textbf{24.22} & 65.99 &21.36      & \textbf{54.36} 
		& 49.82  & \textbf{62.96} & 47.06  & 68.79    & \textbf{56.79}    \\ \bottomrule
	\end{tabular}
\end{table*}

\subsection{Ablation Study}
To comprehensively analyze the actual contribution of each module of DGDA to the overall model performance, we designed and conducted multiple sets of ablation experiments as shown in Tables \ref{tab:aba1}, \ref{tab:aba2}, \ref{tab:aba3}, and \ref{tab:aba4}.  Specifically, we constructed the following seven variant configurations: (1) DGDA-HGNN: HGNN is used in both branches for graph semantic feature extraction; (2) DGDA-PathNN: PathNN is used in both branches. (3) DGDA/$\sigma^{HGNN}$: The perturbation module is removed from the HGNN branch of DGDA. (4) DGDA/$\sigma^{PathNN}$: The perturbation module is removed from the PathNN branch. (5) DGDA/AP: The perturbation module is removed from both branches. (6) DGDA/BC: The branch coupling module is removed. (7) DGDA/RT: The regularization term is removed from both branches. DGDA significantly outperforms DGDA-HGNN and DGDA-PathNN in overall performance, indicating that a single graph semantic modeling method is not sufficient to extract and fuse graph semantic features from different perspectives. DGDA shows significant performance advantages in comparison with the three versions of the perturbation removal module (DGDA/$\sigma^{HGNN}$, DGDA/$\sigma^{PathNN}$, and DGDA/AP). By introducing perturbations in the feature space, the model can effectively prevent overfitting of the source domain features. Due to the lack of coupling and category alignment mechanism between branches, DGDA/BC has a significantly insufficient ability to distinguish categories in the target domain. Furthermore, without the introduction of regularization loss, the overall performance of the model shows a certain degree of degradation when facing training data with noisy labels.

\begin{table*}[htbp]
	\centering
	\caption{The performance of different methods is shown under different noisy rates on the IEMOCAP and MELD datasets. The arrow $\rightarrow$ means from source to target domains. \textbf{Bold} results indicate the best performance. We set the noise rate to 40\%.}
	\label{tab:aba4}
	\renewcommand{\arraystretch}{1} 
	\setlength{\tabcolsep}{3pt} 
	\begin{tabular}{>{\arraybackslash}p{2.3cm}>{\centering\arraybackslash}p{1.1cm}>{\centering\arraybackslash}p{1.1cm}>{\centering\arraybackslash}p{1.1cm}>{\centering\arraybackslash}p{1.1cm}>{\centering\arraybackslash}p{1.1cm}>{\centering\arraybackslash}p{1.1cm}>{\centering\arraybackslash}p{1.1cm}>{\centering\arraybackslash}p{1.1cm}>{\centering\arraybackslash}p{1.1cm}>{\centering\arraybackslash}p{1.1cm}>{\centering\arraybackslash}p{1.1cm}}
		\toprule
		\multirow{2}{*}{Methods} & \multicolumn{5}{c}{IEMOCAP $\rightarrow$ MELD} & \multicolumn{5}{c}{MELD $\rightarrow$ IEMOCAP} \\ \cline{2-11} 
		& Joy   & Sadness & Neutral & Anger  & WF1   & Joy   & Sadness & Neutral & Anger & WF1   \\ \midrule
		DGDA-HGNN                  &29.05 & 2.17  & 60.09 & 22.28 & 43.37 & 37.82 & 44.14 & 35.26 & 44.37 & 40.05 \\
		DGDA-PathNN                    &33.60 & 4.32  & 51.39 & 19.75 & 39.23 & 33.53 & 36.77 & 38.02 & 48.31 & 39.77 \\
		DGDA/$\sigma^{HGNN}$             &33.23 & 6.86  & 64.70 & 6.45  & 45.07 & 31.58 & 39.01 & 34.45 & \textbf{61.35} & 42.08 \\
		DGDA/$\sigma^{PathNN}$                    &35.08 & 3.19  & 60.37 & 8.89  & 43.02 & \textbf{38.82} & 31.96 & 40.72 & 46.25 & 39.85 \\
		DGDA/AP                  &17.75 & 1.72  & 56.67 & 16.23 & 38.20 & 26.81 & 31.16 & \textbf{47.53} & 48.98 & 41.19 \\
		DGDA/BC    & 11.91 & 2.98  & 47.58 & 17.50 & 32.12 & 23.35 & 43.63 & 36.36 & 38.76 & 36.85 \\
		DGDA/RT    & \textbf{38.22} & 3.21  & 59.93 & 25.67 & 45.75 & 26.23 & 37.51 & 32.33 & 47.81 & 36.70  \\ \midrule
		\rowcolor{gray!20}
		DGDA & 37.85 & \textbf{13.51}  &\textbf{64.74}  &\textbf{28.91}    & \textbf{49.74}  
		&35.58 & \textbf{56.90}  & 35.51  & 57.51  & \textbf{46.21}     \\ \bottomrule
	\end{tabular}
\end{table*}

\subsection{Effect of Different Modalities}
\begin{figure}[htbp]
	\centering
	\includegraphics[width=0.6\textwidth]{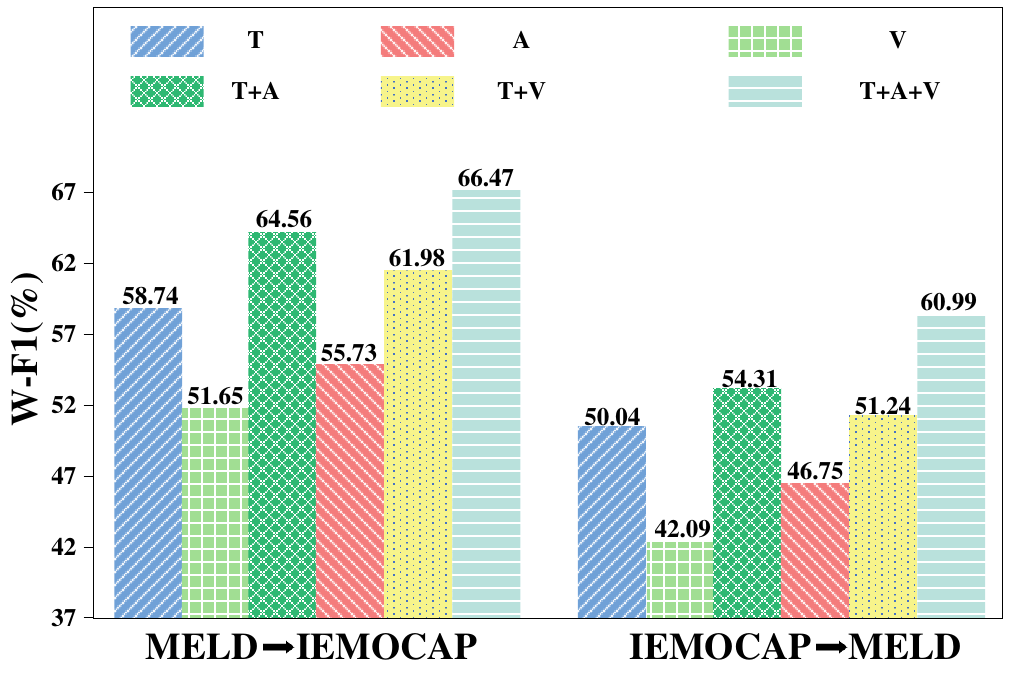} 
	\caption{Verify the effectiveness of multimodal features.}
	\label{tab:multimodal}
\end{figure}

To further explore the contribution of different modalities in emotion recognition, we designed a modality ablation experiment under the experimental condition of 10\% noise rate. By gradually removing one or two modalities, we observed the performance differences of the model under various combinations. Table \ref{tab:multimodal} shows the results of different modal combinations. First, for the single-modal experimental results, the performance of the text modality is far better than that of the audio modality and the visual modality. This phenomenon shows that although multimodal information has a synergistic effect, text features are still the most critical basis for emotion discrimination. Secondly, in the dual-modal combination experiment, the performance of all combinations is better than the corresponding single-modal results, which verifies that there is a complementary relationship between different modalities and joint modeling helps to improve the emotion recognition effect. Finally, when the three modalities are simultaneously involved in feature modeling, the model achieves the best recognition effect, significantly better than all single-modal and bimodal configurations.

\begin{figure*}[htbp]
	\centering
	\begin{subfigure}[b]{0.48\linewidth}
		\centering
		\includegraphics[width=\textwidth]{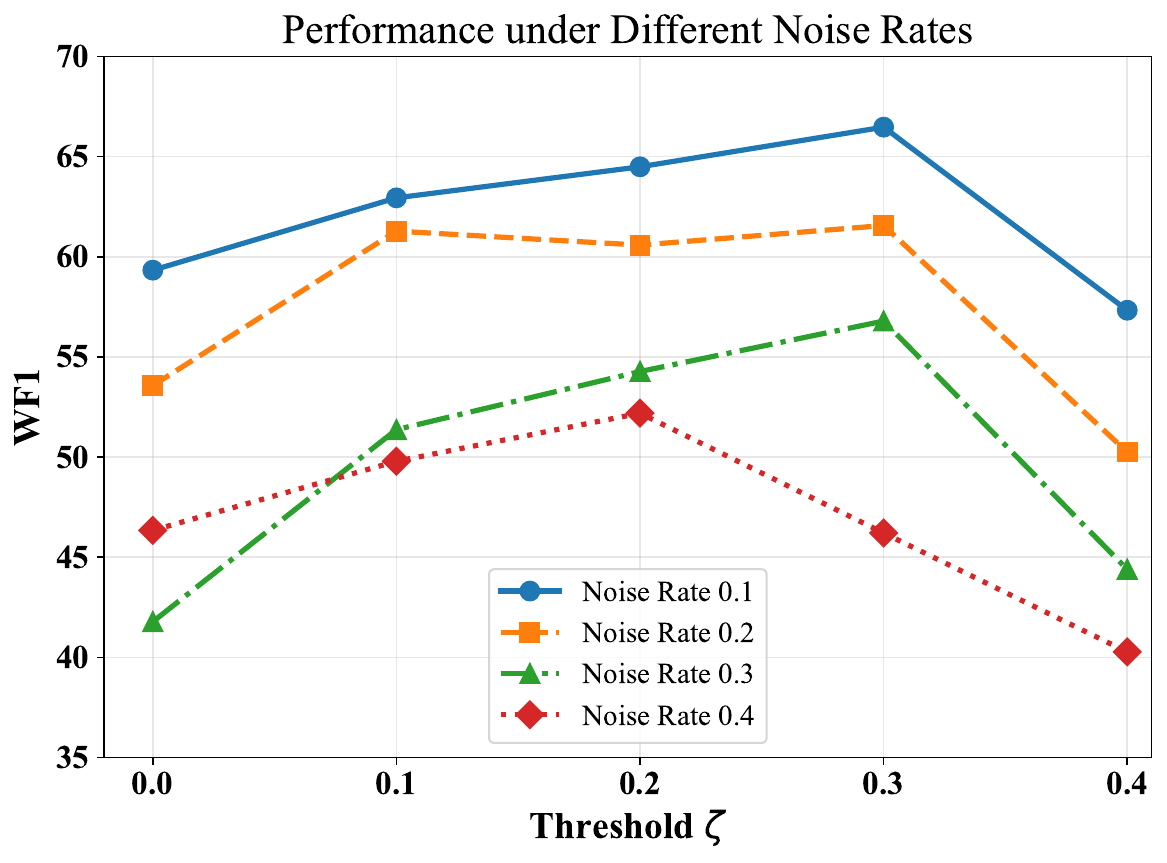}
		\caption{MELD $\rightarrow$ IEMOCAP}
		\label{fig:embed_visual_emo_initial_iemocap6}
	\end{subfigure}
	\hfill
	\begin{subfigure}[b]{0.48\linewidth}
		\centering
		\includegraphics[width=\textwidth]{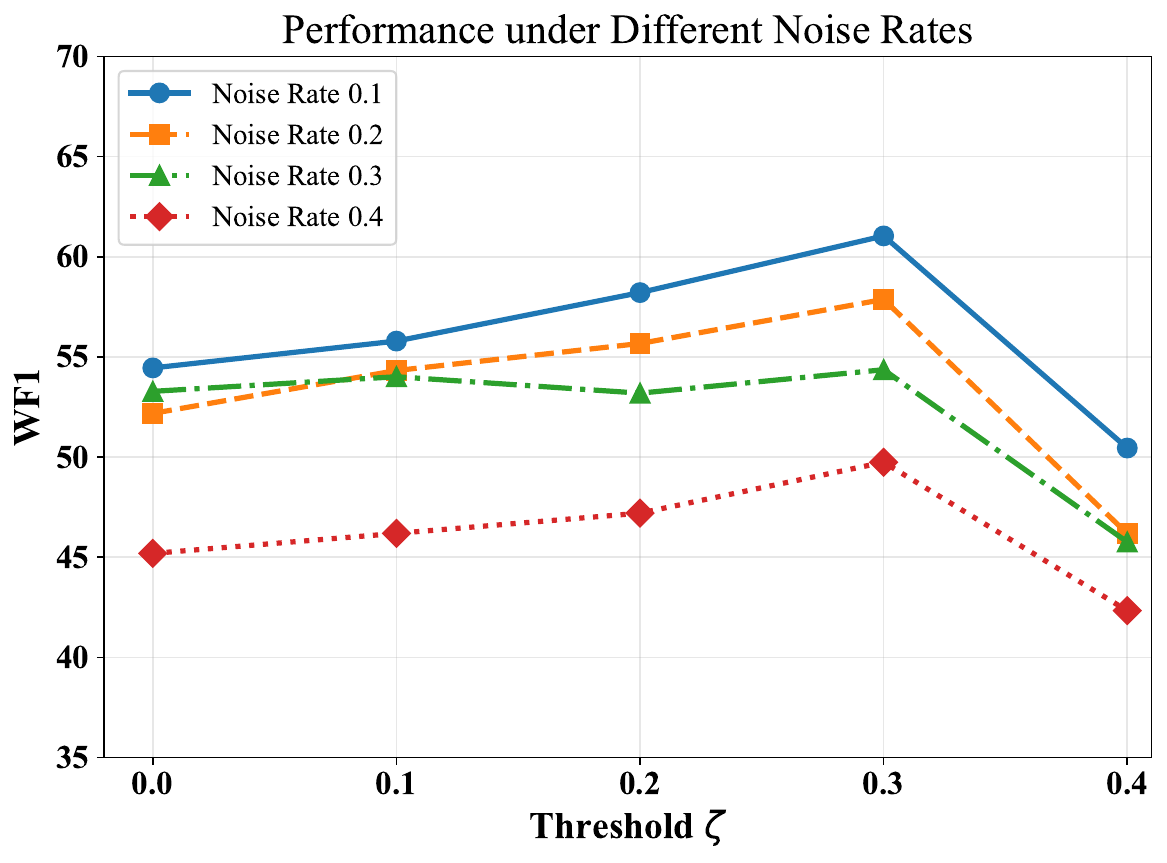}
		\caption{IEMOCAP $\rightarrow$ MELD}
		\label{fig:embed_visual_emo_mmgcn_iemocap6}
	\end{subfigure}
	\vspace{1.5em} 
	\begin{subfigure}[b]{0.48\linewidth}
		\centering
		\includegraphics[width=\textwidth]{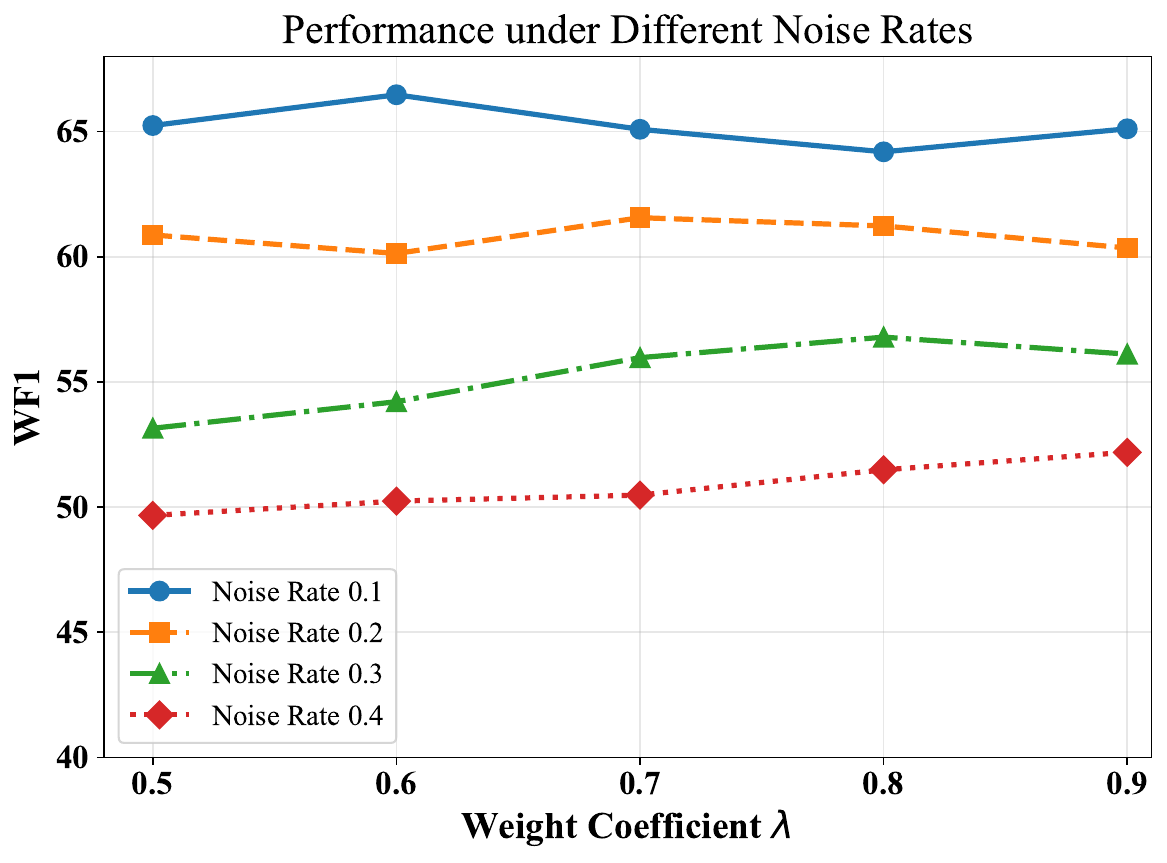}
		\caption{MELD $\rightarrow$ IEMOCAP}
		\label{fig:embed_visual_emo_m3net_iemocap6}
	\end{subfigure}
	\hfill
	\begin{subfigure}[b]{0.48\linewidth}
		\centering
		\includegraphics[width=\textwidth]{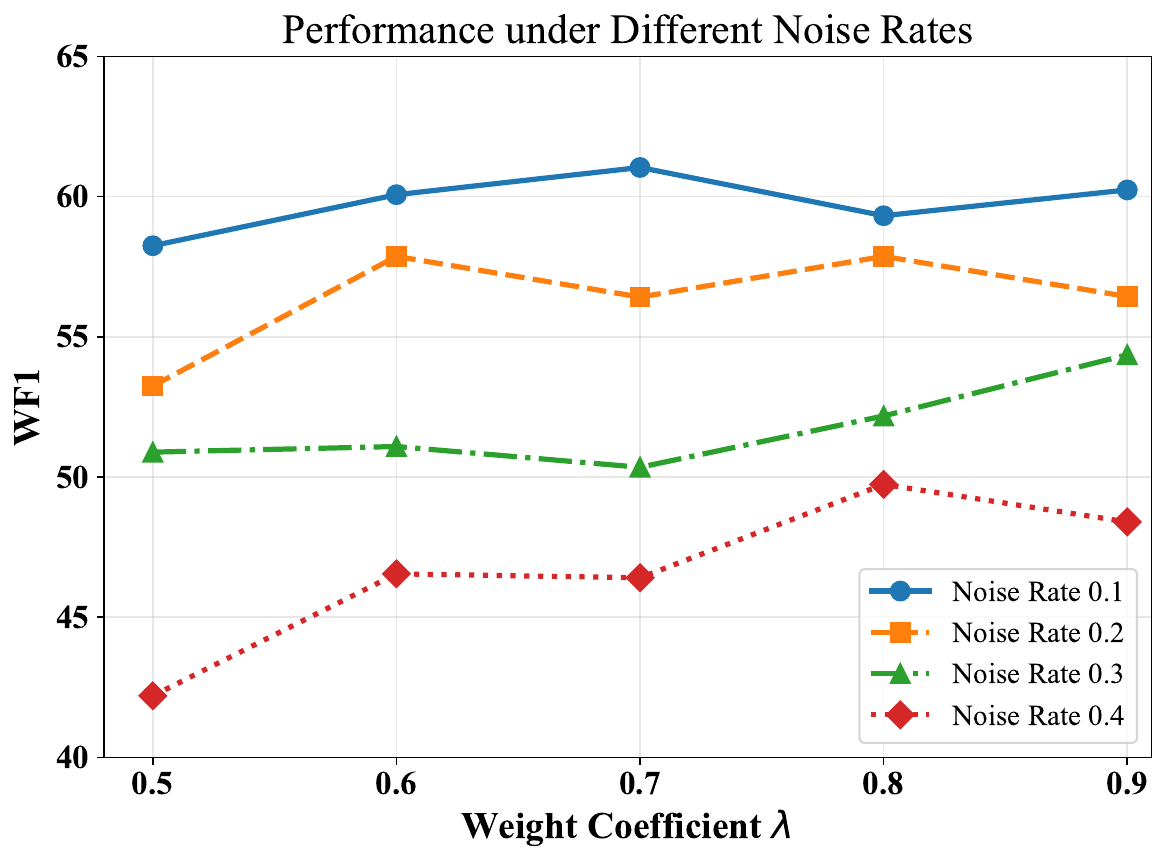}
		\caption{IEMOCAP $\rightarrow$ MELD}
		\label{fig:embed_visual_emo_graphsmile_iemocap6}
	\end{subfigure}
	\caption{Hyperparameter sensitivity of threshold $\zeta$ and regularization weight $\lambda$.}
	\label{fig:sen}
\end{figure*}

\begin{figure*}[htbp]
	\centering
	\begin{subfigure}[b]{0.24\linewidth}
		\centering
		\includegraphics[width=\textwidth]{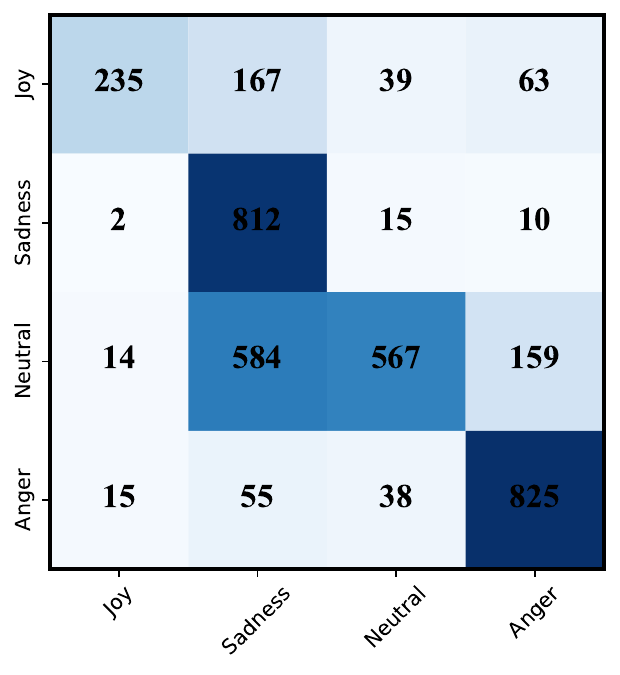}
		\caption{10\% noise}
		\label{fig:embed_visual_emo_initial_iemocap6}
	\end{subfigure}
	\hfill
	\begin{subfigure}[b]{0.24\linewidth}
		\centering
		\includegraphics[width=\textwidth]{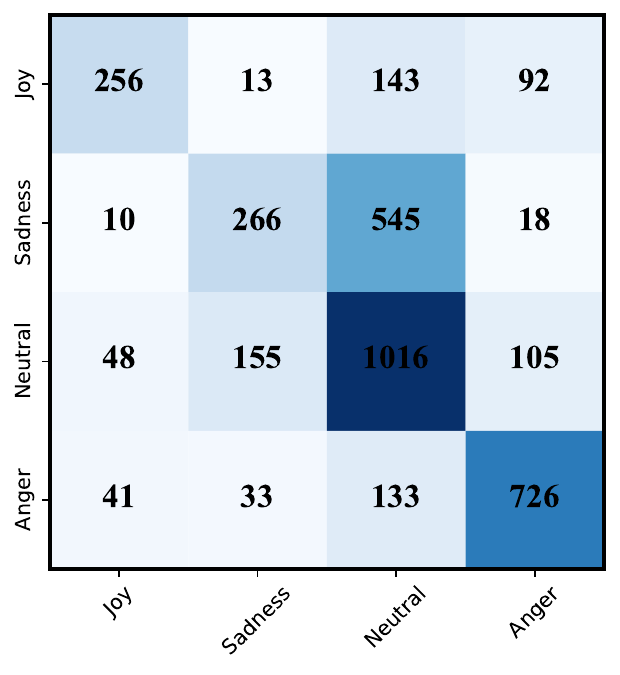}
		\caption{20\% noise}
		\label{fig:embed_visual_emo_mmgcn_iemocap6}
	\end{subfigure}
	\hfill
	\begin{subfigure}[b]{0.24\linewidth}
		\centering
		\includegraphics[width=\textwidth]{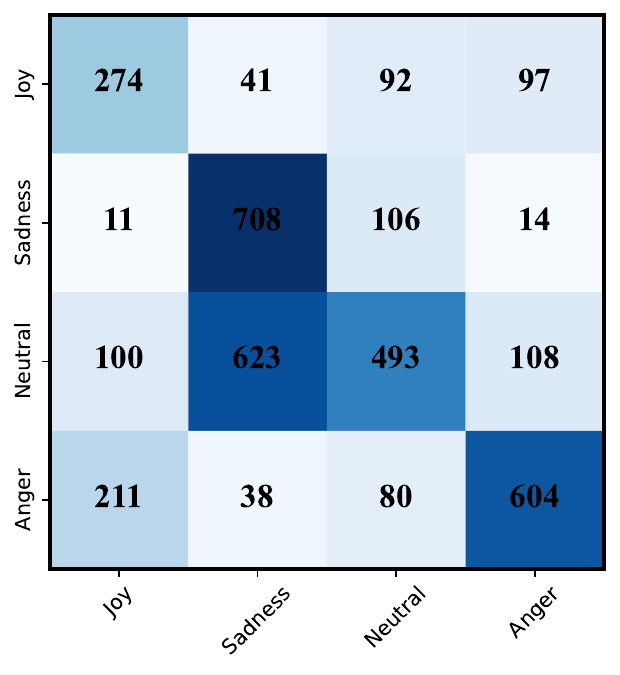}
		\caption{30\% noise}
		\label{fig:embed_visual_emo_m3net_iemocap6}
	\end{subfigure}
	\hfill
	\begin{subfigure}[b]{0.24\linewidth}
		\centering
		\includegraphics[width=\textwidth]{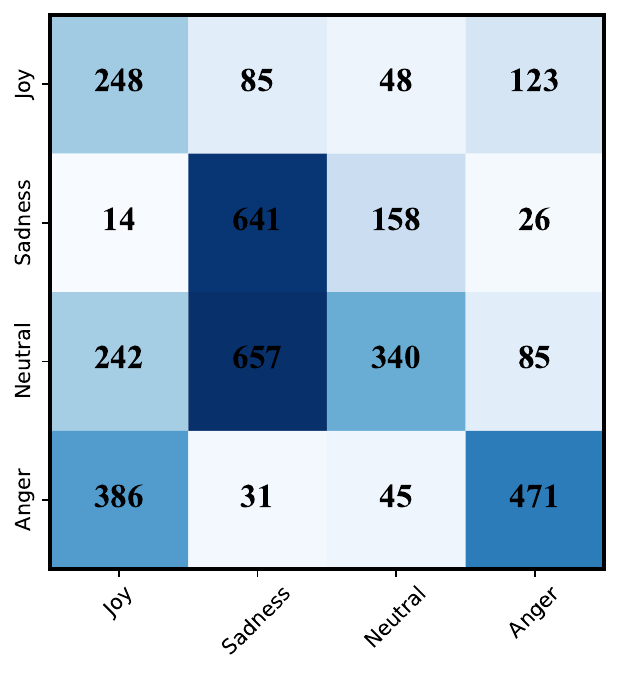}
		\caption{40\% noise}
		\label{fig:embed_visual_emo_graphsmile_iemocap6}
	\end{subfigure}
	\vspace{1.5em} 
	\begin{subfigure}[b]{0.24\linewidth}
		\centering
		\includegraphics[width=\textwidth]{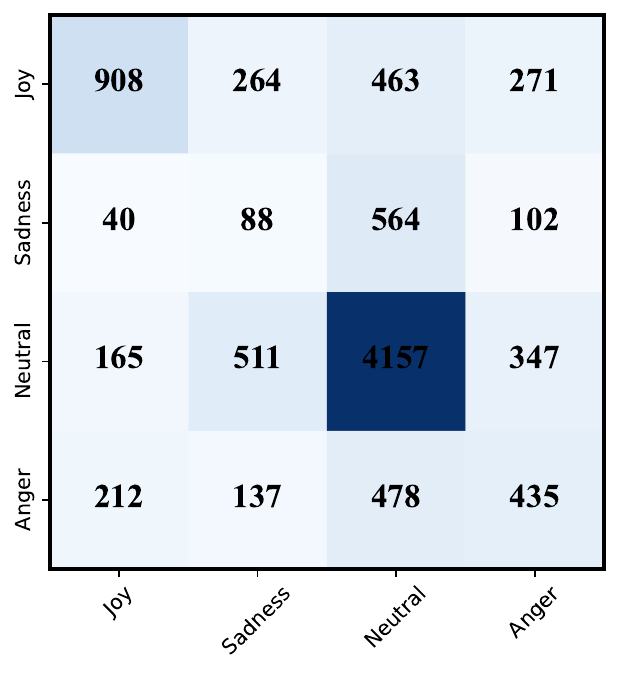}
		\caption{10\% noise}
		\label{fig:embed_visual_emo_initial_meld}
	\end{subfigure}
	\hfill
	\begin{subfigure}[b]{0.24\linewidth}
		\centering
		\includegraphics[width=\textwidth]{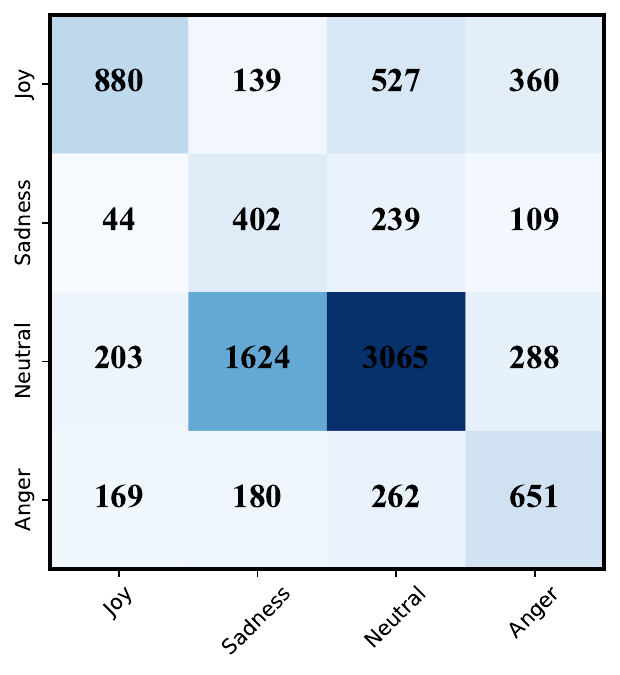}
		\caption{20\% noise}
		\label{fig:embed_visual_emo_mmgcn_meld}
	\end{subfigure}
	\hfill
	\begin{subfigure}[b]{0.24\linewidth}
		\centering
		\includegraphics[width=\textwidth]{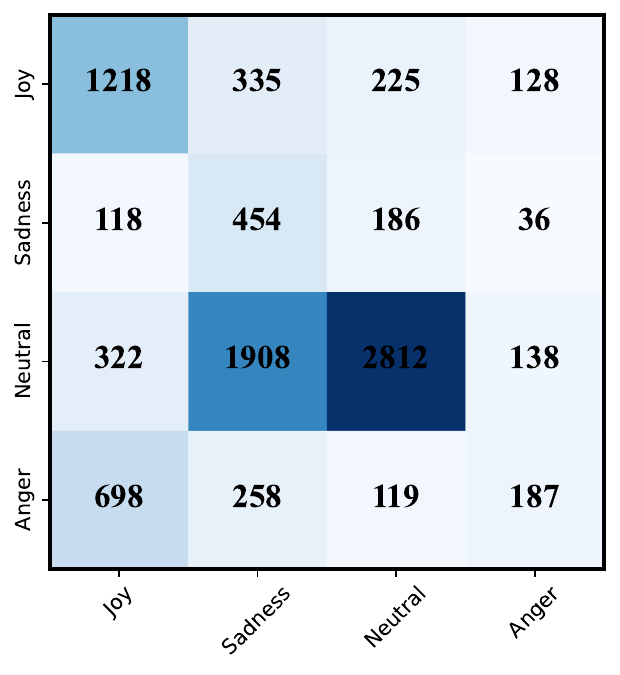}
		\caption{30\% noise}
		\label{fig:embed_visual_emo_m3net_meld}
	\end{subfigure}
	\hfill
	\begin{subfigure}[b]{0.24\linewidth}
		\centering
		\includegraphics[width=\textwidth]{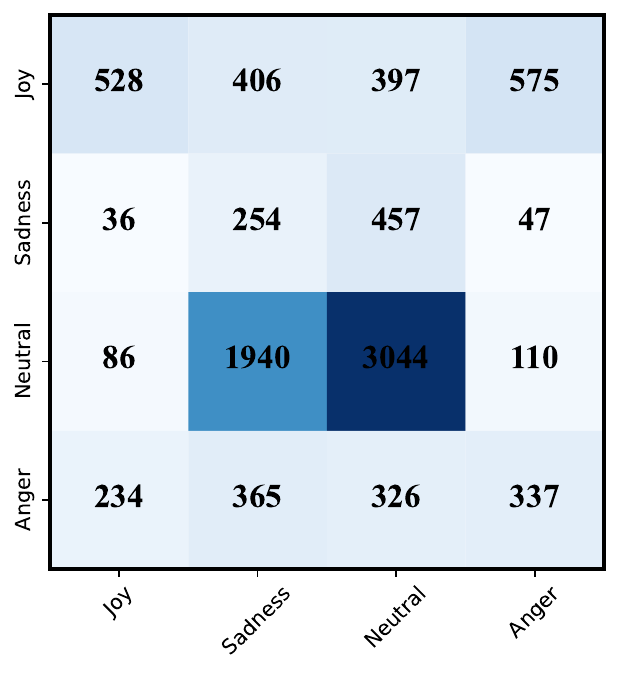}
		\caption{40\% noise}
		\label{fig:embed_visual_emo_graphsmile_meld}
	\end{subfigure}
	\caption{Confusion matrices for multimodal emotion recognition datasets. The matrices provide insights into the model’s classification accuracy, highlighting the challenges and successes in distinguishing between different emotional categories. \textbf{Top:} Results with varying noise levels on the first dataset setting. \textbf{Bottom:} Results with varying noise levels on the second dataset setting.}
	\label{fig:confuse}
\end{figure*}

\subsection{Sensitivity Analysis}

Figs. \ref{fig:sen} (a) and (b) show the effect of the pseudo-label selection threshold $\zeta$ on model performance under different noise rates. When the noise rate is 0.1, as the threshold increases, the performance of the model increases significantly, and reaches the optimal value at 0.3, and then decreases slightly when the threshold is too high. This shows that under low-noise conditions, appropriately improving the pseudo-label selection criteria can help screen out more reliable pseudo-labels. However, when the threshold is further increased, the number of optional pseudo-label samples decreases significantly, resulting in insufficient supervision signals. Under high noise rates, the impact of threshold changes on model performance is relatively gentle. This is mainly because in a high-noise environment, the quality of the candidate pseudo-labels themselves is poor, and it is difficult to completely solve the pseudo-label noise problem by simply increasing the threshold. Figs. \ref{fig:sen} (c) and (d) further analyze the impact of the regularization weight $\lambda$ on model performance. The overall trend shows that at lower $\lambda$, the model performance is poor, especially at high noise rates, and the WFI performance is low. As $\lambda$ gradually increases, the effect of the model at each noise rate is generally improved, and the best state is reached when $\lambda$ is about 0.7$\sim$0.8. This shows that appropriately increasing the weight of the regularization term can better suppress the overfitting of the model to the noisy pseudo-labels, especially in a high-noise environment. However, it is worth noting that when $\lambda$ continues to increase to 0.9, the model performance under some noise rates decreases slightly.

\begin{figure*}[htbp]
	\centering
	\begin{subfigure}[b]{0.24\linewidth}
		\centering
		\includegraphics[width=\textwidth]{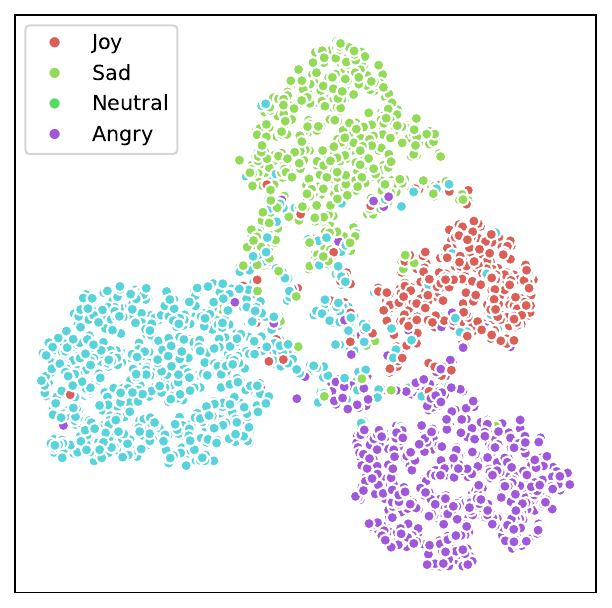}
		\caption{10\% noise}
		\label{fig:embed_visual_emo_initial_iemocap6}
	\end{subfigure}
	\hfill
	\begin{subfigure}[b]{0.24\linewidth}
		\centering
		\includegraphics[width=\textwidth]{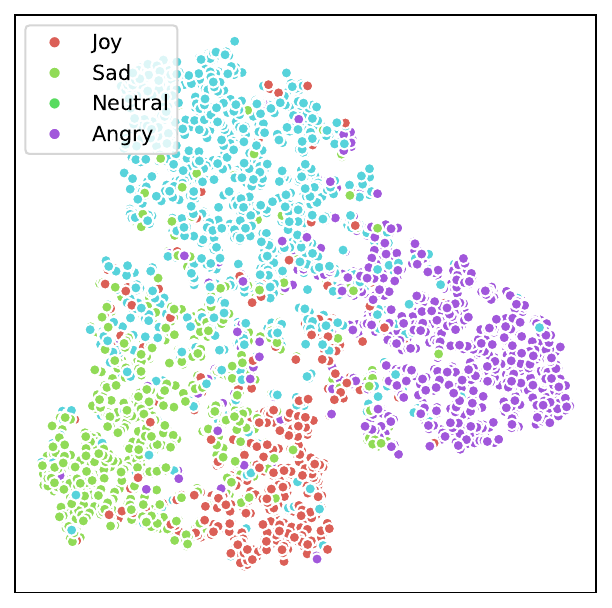}
		\caption{20\% noise}
		\label{fig:embed_visual_emo_mmgcn_iemocap6}
	\end{subfigure}
	\hfill
	\begin{subfigure}[b]{0.24\linewidth}
		\centering
		\includegraphics[width=\textwidth]{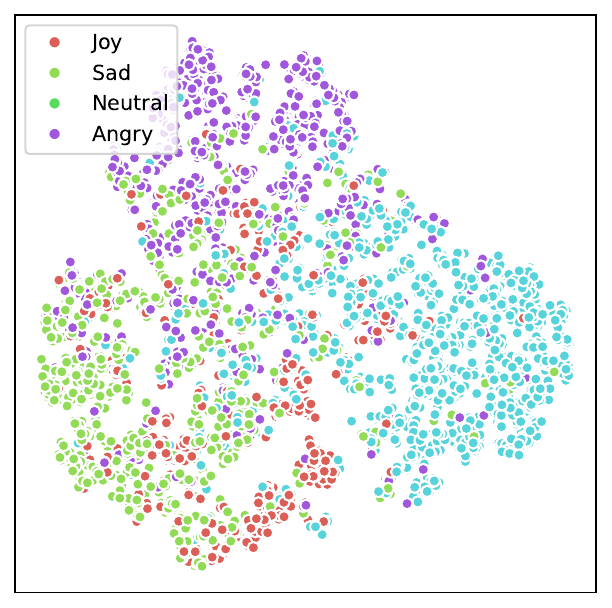}
		\caption{30\% noise}
		\label{fig:embed_visual_emo_m3net_iemocap6}
	\end{subfigure}
	\hfill
	\begin{subfigure}[b]{0.24\linewidth}
		\centering
		\includegraphics[width=\textwidth]{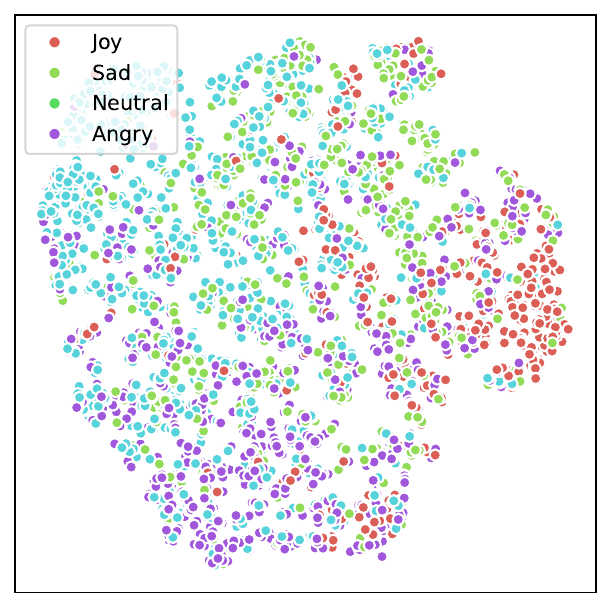}
		\caption{40\% noise}
		\label{fig:embed_visual_emo_graphsmile_iemocap6}
	\end{subfigure}
	\vspace{2em} 
	\begin{subfigure}[b]{0.24\linewidth}
		\centering
		\includegraphics[width=\textwidth]{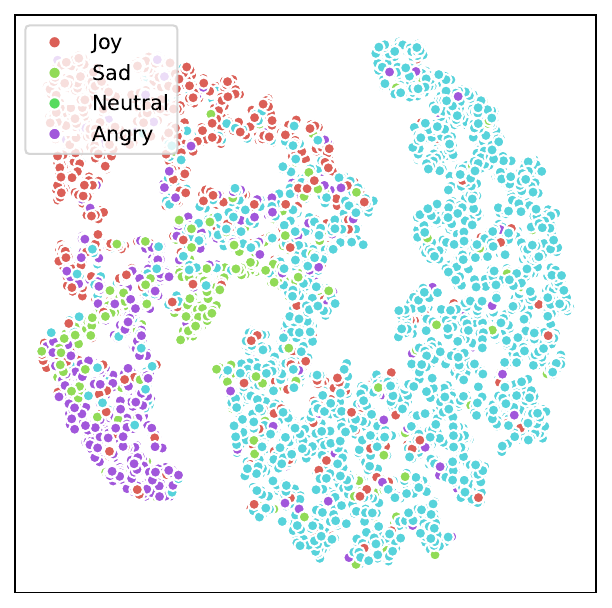}
		\caption{10\% noise}
		\label{fig:embed_visual_emo_initial_meld}
	\end{subfigure}
	\hfill
	\begin{subfigure}[b]{0.24\linewidth}
		\centering
		\includegraphics[width=\textwidth]{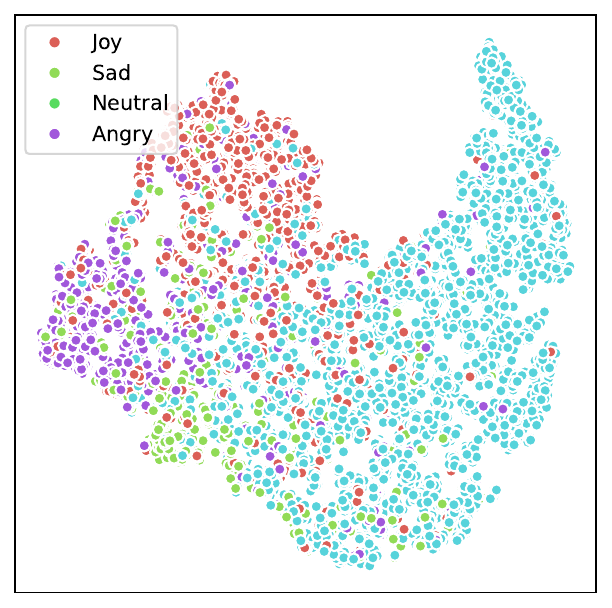}
		\caption{20\% noise}
		\label{fig:embed_visual_emo_mmgcn_meld}
	\end{subfigure}
	\hfill
	\begin{subfigure}[b]{0.24\linewidth}
		\centering
		\includegraphics[width=\textwidth]{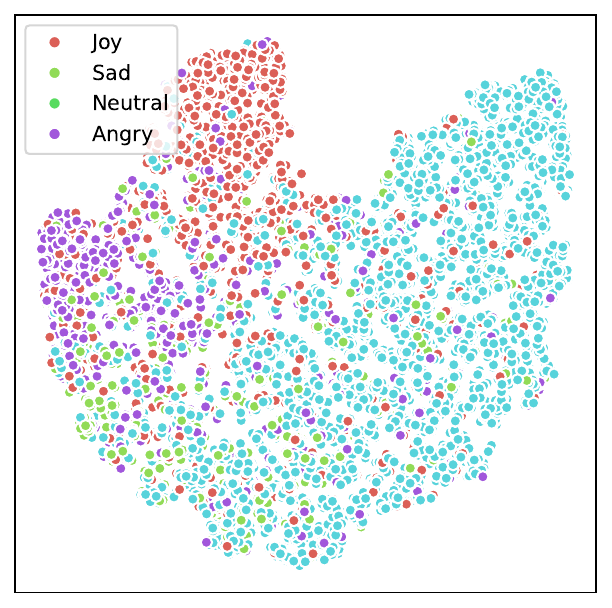}
		\caption{30\% noise}
		\label{fig:embed_visual_emo_m3net_meld}
	\end{subfigure}
	\hfill
	\begin{subfigure}[b]{0.24\linewidth}
		\centering
		\includegraphics[width=\textwidth]{picture/\detokenize{DGDA_MELD-noise-0.4_.pdf}}
		\caption{40\% noise}
		\label{fig:embed_visual_emo_graphsmile_meld}
	\end{subfigure}
	\caption{Visualization of the learned embeddings. \textbf{Top:} Trained on the MELD dataset and tested on the IEMOCAP dataset (with varying noise levels). \textbf{Bottom:} Trained on the IEMOCAP dataset and tested on the MELD dataset (with varying noise levels).}
	\label{fig:figure5}
\end{figure*}

\subsection{Confusion Matrices}
\label{sec:confusem}

Fig. \ref{fig:confuse} shows the confusion matrix of the model prediction results on the IEMOCAP and MELD datasets, which provides an important basis for an in-depth understanding of the model's classification ability and performance differences in different emotion categories. It can be observed that the model has a relatively ideal recognition effect on the two categories of ``Joy" and ``Neutral". Most of the samples belonging to these two categories are accurately classified, and the misclassification ratio between the two is low. This phenomenon shows that the model has successfully learned the discriminative features that are highly related to these two categories of emotions and can effectively distinguish them, reflecting its good modeling ability for common emotion categories. However, the confusion matrix also reveals the classification difficulties in low-frequency emotion categories such as ``Sadness" and ``Anger". Compared with common emotions such as ``Joy" and ``Neutral", these categories already have the problem of insufficient sample number in multimodal emotion recognition datasets, and the category imbalance phenomenon is more significant. Due to the small number of ``Sadness" and ``Anger" samples available for learning during the training process, the model has certain difficulties in capturing the key patterns and features related to these two categories of emotions, resulting in its insufficient generalization ability on these two categories of samples. Therefore, the model's recognition effect on these two types of emotions is significantly inferior, and a high proportion of misclassification occurs. Specifically, the confusion matrix shows that a considerable number of samples in the ``Sadness" and ``Anger" categories are mistakenly classified into other categories such as ``Happy" or ``Neutral". This result reflects that the model has a certain degree of discrimination ambiguity between these emotions; that is, when distinguishing low-frequency emotion categories from common emotion categories, it may rely on overlapping or fuzzy features in emotional expressions, which leads to confusion. 


\subsection{Visualization}

Fig. \ref{fig:figure5} shows the distribution of emotional features learned by the model on the IEMOCAP and MELD datasets under different noise ratios (10\%, 20\%, 30\%, and 40\%), which intuitively reveals the impact of noise level changes on the model's discriminative ability and the effectiveness of the proposed method in combating noise interference. From the visualization results, on the IEMOCAP dataset (upper row in the figure), as the noise ratio gradually increases, the boundaries between different emotional categories begin to become blurred, the sample distribution within the category tends to be loose, and the feature overlap between categories becomes more obvious. Especially when the noise ratio reaches 30\% and 40\%, the degree of confusion between emotional categories increases significantly, and samples of categories such as Sad and Neutral, Joy and Angry appear to be distributed in a large area, resulting in a sharp decline in the model's ability to distinguish these emotional categories, and the reliability of the discrimination results is greatly affected. This phenomenon reflects that in a medium-to-high noise environment, the potential feature commonality between emotional categories and the perturbation effect of noise increase the learning difficulty of the model, making it difficult to maintain the effective extraction of discriminative features. In contrast, the MELD dataset (lower line in the figure) shows stronger robustness and category separability under the same noise level. Although the noise ratio also increased from 10\% to 40\%, the feature distribution between emotion categories still maintained a good clustering structure and relatively clear category boundaries. Even under the most stringent 40\% noise condition, the samples of each category still showed a relatively stable distribution pattern, and the distinction between categories was preserved to a certain extent. This result fully demonstrates that the proposed method exhibits superior noise resistance on the MELD dataset, and can effectively suppress the erosion of the model feature space by erroneous labels in an environment with significant noise interference, ensuring that the model maintains good discrimination ability for emotion categories. It is worth noting that the MELD dataset itself is more challenging in terms of modal combination, context diversity, and speaker complexity. Therefore, the model can still show strong noise resistance on this dataset, which further proves the robustness and generalization ability of the proposed method in the face of complex multimodal data and label uncertainty.

\section{Conclusions}

In this paper, we propose a Dual-branch Graph Domain Adaptation (DGDA) for multi-modal emotion recognition in cross-scenario conversations. Specifically, we first construct an emotion interaction graph to model the complex emotional dependencies between utterances. Then, we design a neighborhood aggregation and path aggregation dual-branch graph encoder to explicitly and implicitly capture the dynamic changes in emotion between utterances and explore multivariate relationships, respectively. To address the problem of out-of-domain distribution differences, we introduce a domain adversarial classifier to improve the representation ability of invariant features in the source domain. Furthermore, we construct a regularization loss to prevent the model from memorizing noise and improve the model's ability to resist interference from noisy labels.	

\bibliography{pami}
\bibliographystyle{tmlr}

\end{document}

%% file: math_commands.tex
\usepackage{amsmath,amsfonts,bm}

\def\eqref#1{equation~\ref{#1}}

\def\1{\bm{1}}

\DeclareMathAlphabet{\mathsfit}{\encodingdefault}{\sfdefault}{m}{sl}
\SetMathAlphabet{\mathsfit}{bold}{\encodingdefault}{\sfdefault}{bx}{n}













%% file: main2.bbl
\begin{thebibliography}{96}
\providecommand{\natexlab}[1]{#1}
\providecommand{\url}[1]{\texttt{#1}}
\expandafter\ifx\csname urlstyle\endcsname\relax
  \providecommand{\doi}[1]{doi: #1}\else
  \providecommand{\doi}{doi: \begingroup \urlstyle{rm}\Url}\fi

\bibitem[Ai et~al.(2024)Ai, Shou, Meng, and Li]{ai2024gcn}
Wei Ai, Yuntao Shou, Tao Meng, and Keqin Li.
\newblock Der-gcn: Dialog and event relation-aware graph convolutional neural
  network for multimodal dialog emotion recognition.
\newblock \emph{IEEE Transactions on Neural Networks and Learning Systems},
  2024.

\bibitem[Ai et~al.(2025)Ai, Zhang, Shou, Meng, Chen, and Li]{ai2025revisiting}
Wei Ai, Fuchen Zhang, Yuntao Shou, Tao Meng, Haowen Chen, and Keqin Li.
\newblock Revisiting multimodal emotion recognition in conversation from the
  perspective of graph spectrum.
\newblock In \emph{Proceedings of the AAAI Conference on Artificial
  Intelligence}, volume~39, pp.\  11418--11426, 2025.

\bibitem[Ai et~al.(2026)Ai, Tan, Shou, Meng, Chen, He, and Li]{ai2026paradigm}
Wei Ai, Yilong Tan, Yuntao Shou, Tao Meng, Haowen Chen, Zhixiong He, and Keqin
  Li.
\newblock The paradigm shift: A comprehensive survey on large vision language
  models for multimodal fake news detection.
\newblock \emph{Computer Science Review}, 60:\penalty0 100893, 2026.

\bibitem[Blitzer et~al.(2007)Blitzer, Crammer, Kulesza, Pereira, and
  Wortman]{blitzer2007learning}
John Blitzer, Koby Crammer, Alex Kulesza, Fernando Pereira, and Jennifer
  Wortman.
\newblock Learning bounds for domain adaptation.
\newblock \emph{Advances in neural information processing systems}, 20, 2007.

\bibitem[Busso et~al.(2008)Busso, Bulut, Lee, Kazemzadeh, Mower, Kim, Chang,
  Lee, and Narayanan]{busso2008iemocap}
Carlos Busso, Murtaza Bulut, Chi-Chun Lee, Abe Kazemzadeh, Emily Mower, Samuel
  Kim, Jeannette~N Chang, Sungbok Lee, and Shrikanth~S Narayanan.
\newblock Iemocap: Interactive emotional dyadic motion capture database.
\newblock \emph{Language resources and evaluation}, 42:\penalty0 335--359,
  2008.

\bibitem[Chen et~al.(2024)Chen, Li, Kou, Du, Li, Wang, and
  Vong]{chen2024comprehensive}
Chuangquan Chen, Zhencheng Li, Kit~Ian Kou, Jie Du, Chen Li, Hongtao Wang, and
  Chi-Man Vong.
\newblock Comprehensive multisource learning network for cross-subject
  multimodal emotion recognition.
\newblock \emph{IEEE Transactions on Emerging Topics in Computational
  Intelligence}, 2024.

\bibitem[Chen et~al.(2023)Chen, Shao, Zhu, and Shen]{chen2023multivariate}
Feiyu Chen, Jie Shao, Shuyuan Zhu, and Heng~Tao Shen.
\newblock Multivariate, multi-frequency and multimodal: Rethinking graph neural
  networks for emotion recognition in conversation.
\newblock In \emph{Proceedings of the IEEE/CVF Conference on Computer Vision
  and Pattern Recognition}, pp.\  10761--10770, 2023.

\bibitem[Chen et~al.(2025)Chen, Ye, Wang, Zhang, Zhang, Wang, Zhang, and
  Zhuang]{chen2025smoothness}
Wei Chen, Guo Ye, Yakun Wang, Zhao Zhang, Libang Zhang, Daixin Wang, Zhiqiang
  Zhang, and Fuzhen Zhuang.
\newblock Smoothness really matters: A simple yet effective approach for
  unsupervised graph domain adaptation.
\newblock In \emph{Proceedings of the AAAI Conference on Artificial
  Intelligence}, volume~39, pp.\  15875--15883, 2025.

\bibitem[Dan et~al.(2024)Dan, Liu, Xie, Yu, Dong, and Tan]{dan2024tfgda}
Jun Dan, Weiming Liu, Xie Xie, Hua Yu, Shunjie Dong, and Yanchao Tan.
\newblock Tfgda: Exploring topology and feature alignment in semi-supervised
  graph domain adaptation through robust clustering.
\newblock \emph{Advances in Neural Information Processing Systems},
  37:\penalty0 50230--50255, 2024.

\bibitem[Eyben et~al.(2010)Eyben, W{\"o}llmer, and
  Schuller]{eyben2010opensmile}
Florian Eyben, Martin W{\"o}llmer, and Bj{\"o}rn Schuller.
\newblock Opensmile: the munich versatile and fast open-source audio feature
  extractor.
\newblock In \emph{Proceedings of the 18th ACM international conference on
  Multimedia}, pp.\  1459--1462, 2010.

\bibitem[Ghosh et~al.(2017)Ghosh, Chollet, Laksana, Morency, and
  Scherer]{ghosh2017affect}
Sayan Ghosh, Mathieu Chollet, Eugene Laksana, Louis-Philippe Morency, and
  Stefan Scherer.
\newblock Affect-lm: A neural language model for customizable affective text
  generation.
\newblock \emph{arXiv preprint arXiv:1704.06851}, 2017.

\bibitem[Guo et~al.(2025)Guo, Jin, Xu, Lin, and Wu]{guo2025bridging}
Zirun Guo, Tao Jin, Wenlong Xu, Wang Lin, and Yangyang Wu.
\newblock Bridging the gap for test-time multimodal sentiment analysis.
\newblock In \emph{Proceedings of the AAAI Conference on Artificial
  Intelligence}, volume~39, pp.\  16987--16995, 2025.

\bibitem[Hazmoune \& Bougamouza(2024)Hazmoune and
  Bougamouza]{hazmoune2024using}
Samira Hazmoune and Fateh Bougamouza.
\newblock Using transformers for multimodal emotion recognition: Taxonomies and
  state of the art review.
\newblock \emph{Engineering Applications of Artificial Intelligence},
  133:\penalty0 108339, 2024.

\bibitem[Ho et~al.(2020)Ho, Yang, Kim, and Lee]{ho2020multimodal}
Ngoc-Huynh Ho, Hyung-Jeong Yang, Soo-Hyung Kim, and Gueesang Lee.
\newblock Multimodal approach of speech emotion recognition using multi-level
  multi-head fusion attention-based recurrent neural network.
\newblock \emph{IEEE Access}, 8:\penalty0 61672--61686, 2020.

\bibitem[Hu et~al.(2021)Hu, Liu, Zhao, and Jin]{hu2021mmgcn}
Jingwen Hu, Yuchen Liu, Jinming Zhao, and Qin Jin.
\newblock Mmgcn: Multimodal fusion via deep graph convolution network for
  emotion recognition in conversation.
\newblock In \emph{Proceedings of the 59th Annual Meeting of the Association
  for Computational Linguistics and the 11th International Joint Conference on
  Natural Language Processing (Volume 1: Long Papers)}, pp.\  5666--5675, 2021.

\bibitem[Hu et~al.(2024)Hu, Chen, Yang, Qin, Chen, Chng, and Zhang]{hu2024self}
Yuchen Hu, Chen Chen, Chao-Han Yang, Chengwei Qin, Pin-Yu Chen, Eng-Siong Chng,
  and Chao Zhang.
\newblock Self-taught recognizer: Toward unsupervised adaptation for speech
  foundation models.
\newblock \emph{Advances in Neural Information Processing Systems},
  37:\penalty0 29566--29594, 2024.

\bibitem[Huang et~al.(2017)Huang, Liu, Van Der~Maaten, and
  Weinberger]{huang2017densely}
Gao Huang, Zhuang Liu, Laurens Van Der~Maaten, and Kilian~Q Weinberger.
\newblock Densely connected convolutional networks.
\newblock In \emph{Proceedings of the IEEE conference on computer vision and
  pattern recognition}, pp.\  4700--4708, 2017.

\bibitem[Huang et~al.(2020)Huang, Zhu, and Gao]{huang2020challenges}
Minlie Huang, Xiaoyan Zhu, and Jianfeng Gao.
\newblock Challenges in building intelligent open-domain dialog systems.
\newblock \emph{ACM Transactions on Information Systems (TOIS)}, 38\penalty0
  (3):\penalty0 1--32, 2020.

\bibitem[Huddar et~al.(2021)Huddar, Sannakki, and
  Rajpurohit]{huddar2021attention}
Mahesh~G Huddar, Sanjeev~S Sannakki, and Vijay~S Rajpurohit.
\newblock Attention-based multi-modal sentiment analysis and emotion detection
  in conversation using rnn.
\newblock 2021.

\bibitem[Jin et~al.(2024)Jin, Lian, He, and Zhu]{jin2024hgdl}
Yufei Jin, Heng Lian, Yi~He, and Xingquan Zhu.
\newblock Hgdl: Heterogeneous graph label distribution learning.
\newblock \emph{Advances in Neural Information Processing Systems},
  37:\penalty0 40792--40830, 2024.

\bibitem[Kang \& Cho(2025)Kang and Cho]{kang2025beyond}
Yujin Kang and Yoon-Sik Cho.
\newblock Beyond single emotion: Multi-label approach to conversational emotion
  recognition.
\newblock In \emph{Proceedings of the AAAI Conference on Artificial
  Intelligence}, volume~39, pp.\  24321--24329, 2025.

\bibitem[Khare \& Bajaj(2020)Khare and Bajaj]{khare2020time}
Smith~K Khare and Varun Bajaj.
\newblock Time--frequency representation and convolutional neural network-based
  emotion recognition.
\newblock \emph{IEEE transactions on neural networks and learning systems},
  32\penalty0 (7):\penalty0 2901--2909, 2020.

\bibitem[Kim \& Vossen(2021)Kim and Vossen]{kim2021emoberta}
Taewoon Kim and Piek Vossen.
\newblock Emoberta: Speaker-aware emotion recognition in conversation with
  roberta.
\newblock \emph{arXiv preprint arXiv:2108.12009}, 2021.

\bibitem[Kim(2014)]{kim-2014-convolutional}
Yoon Kim.
\newblock Convolutional neural networks for sentence classification.
\newblock In \emph{Proceedings of the 2014 Conference on Empirical Methods in
  Natural Language Processing ({EMNLP})}, pp.\  1746--1751. ACL, October 2014.

\bibitem[Li et~al.(2025)Li, Fei, Li, Wu, Liao, Wei, Chua, and
  Ji]{li2025revisiting}
Bobo Li, Hao Fei, Fei Li, Shengqiong Wu, Lizi Liao, Yinwei Wei, Tat-Seng Chua,
  and Donghong Ji.
\newblock Revisiting conversation discourse for dialogue disentanglement.
\newblock \emph{ACM Transactions on Information Systems}, 43\penalty0
  (1):\penalty0 1--34, 2025.

\bibitem[Li et~al.(2024{\natexlab{a}})Li, Wang, Liu, and Zeng]{li2023cfnesa}
Jiang Li, Xiaoping Wang, Yingjian Liu, and Zhigang Zeng.
\newblock {CFN-ESA}: A cross-modal fusion network with emotion-shift awareness
  for dialogue emotion recognition.
\newblock \emph{IEEE Transactions on Affective Computing}, 15\penalty0
  (4):\penalty0 1919--1933, 2024{\natexlab{a}}.
\newblock \doi{10.1109/TAFFC.2024.3389453}.

\bibitem[Li et~al.(2024{\natexlab{b}})Li, Wang, Liu, and Zeng]{li2024cfn}
Jiang Li, Xiaoping Wang, Yingjian Liu, and Zhigang Zeng.
\newblock Cfn-esa: A cross-modal fusion network with emotion-shift awareness
  for dialogue emotion recognition.
\newblock \emph{IEEE Transactions on Affective Computing}, 2024{\natexlab{b}}.

\bibitem[Li et~al.(2021)Li, Xiong, and Hoi]{li2021learning}
Junnan Li, Caiming Xiong, and Steven~CH Hoi.
\newblock Learning from noisy data with robust representation learning.
\newblock In \emph{Proceedings of the IEEE/CVF International Conference on
  Computer Vision}, pp.\  9485--9494, 2021.

\bibitem[Lian et~al.(2021)Lian, Liu, and Tao]{lian2021ctnet}
Zheng Lian, Bin Liu, and Jianhua Tao.
\newblock Ctnet: Conversational transformer network for emotion recognition.
\newblock \emph{IEEE/ACM Transactions on Audio, Speech, and Language
  Processing}, 29:\penalty0 985--1000, 2021.

\bibitem[Lian et~al.(2023)Lian, Chen, Sun, Liu, and Tao]{lian2023gcnet}
Zheng Lian, Lan Chen, Licai Sun, Bin Liu, and Jianhua Tao.
\newblock Gcnet: Graph completion network for incomplete multimodal learning in
  conversation.
\newblock \emph{IEEE Transactions on pattern analysis and machine
  intelligence}, 45\penalty0 (7):\penalty0 8419--8432, 2023.

\bibitem[Liu et~al.(2024{\natexlab{a}})Liu, Lou, Zhang, Wu, Xiao, Jensen, and
  Zhang]{liu2024eeg}
Huan Liu, Tianyu Lou, Yuzhe Zhang, Yixiao Wu, Yang Xiao, Christian~S Jensen,
  and Dalin Zhang.
\newblock Eeg-based multimodal emotion recognition: a machine learning
  perspective.
\newblock \emph{IEEE Transactions on Instrumentation and Measurement},
  2024{\natexlab{a}}.

\bibitem[Liu et~al.(2024{\natexlab{b}})Liu, Fang, Zhang, Gu, Zhou, Wang, and
  Bu]{liu2024rethinking}
Meihan Liu, Zeyu Fang, Zhen Zhang, Ming Gu, Sheng Zhou, Xin Wang, and Jiajun
  Bu.
\newblock Rethinking propagation for unsupervised graph domain adaptation.
\newblock In \emph{Proceedings of the AAAI Conference on Artificial
  Intelligence}, volume~38, pp.\  13963--13971, 2024{\natexlab{b}}.

\bibitem[Liu et~al.(2020)Liu, Niles-Weed, Razavian, and
  Fernandez-Granda]{liu2020early}
Sheng Liu, Jonathan Niles-Weed, Narges Razavian, and Carlos Fernandez-Granda.
\newblock Early-learning regularization prevents memorization of noisy labels.
\newblock \emph{Advances in neural information processing systems},
  33:\penalty0 20331--20342, 2020.

\bibitem[Liu et~al.(2023)Liu, Li, Wang, and Zeng]{liu2023emotionic}
Yingjian Liu, Jiang Li, Xiaoping Wang, and Zhigang Zeng.
\newblock Emotionic: Emotional inertia and contagion-driven dependency modeling
  for emotion recognition in conversation.
\newblock \emph{SCIENCE CHINA Information Sciences}, 67\penalty0 (8):\penalty0
  182103:1--182103:17, 2023.
\newblock \doi{10.1007/s11432-023-3908-6}.

\bibitem[Lu et~al.(2025)Lu, Chen, Liang, Tan, Zeng, and
  Hu]{lu2025understanding}
Haifeng Lu, Jiuyi Chen, Feng Liang, Mingkui Tan, Runhao Zeng, and Xiping Hu.
\newblock Understanding emotional body expressions via large language models.
\newblock In \emph{Proceedings of the AAAI Conference on Artificial
  Intelligence}, volume~39, pp.\  1447--1455, 2025.

\bibitem[Ma et~al.(2023)Ma, Wang, Lin, Zhang, Zhang, and Xu]{ma2023transformer}
Hui Ma, Jian Wang, Hongfei Lin, Bo~Zhang, Yijia Zhang, and Bo~Xu.
\newblock A transformer-based model with self-distillation for multimodal
  emotion recognition in conversations.
\newblock \emph{IEEE Transactions on Multimedia}, 2023.

\bibitem[Ma et~al.(2024)Ma, Wang, Lin, Zhang, Zhang, and Xu]{10109845}
Hui Ma, Jian Wang, Hongfei Lin, Bo~Zhang, Yijia Zhang, and Bo~Xu.
\newblock A transformer-based model with self-distillation for multimodal
  emotion recognition in conversations.
\newblock \emph{IEEE Transactions on Multimedia}, 26:\penalty0 776--788, 2024.

\bibitem[Majumder et~al.(2019)Majumder, Poria, Hazarika, Mihalcea, Gelbukh, and
  Cambria]{majumder2019dialoguernn}
Navonil Majumder, Soujanya Poria, Devamanyu Hazarika, Rada Mihalcea, Alexander
  Gelbukh, and Erik Cambria.
\newblock Dialoguernn: An attentive rnn for emotion detection in conversations.
\newblock In \emph{Proceedings of the AAAI conference on artificial
  intelligence}, volume~33, pp.\  6818--6825, 2019.

\bibitem[Meng et~al.(2024{\natexlab{a}})Meng, Shou, Ai, Du, Liu, and
  Li]{meng2024multi}
Tao Meng, Yuntao Shou, Wei Ai, Jiayi Du, Haiyan Liu, and Keqin Li.
\newblock A multi-message passing framework based on heterogeneous graphs in
  conversational emotion recognition.
\newblock \emph{Neurocomputing}, 569:\penalty0 127109, 2024{\natexlab{a}}.

\bibitem[Meng et~al.(2024{\natexlab{b}})Meng, Shou, Ai, Yin, and
  Li]{meng2024deep}
Tao Meng, Yuntao Shou, Wei Ai, Nan Yin, and Keqin Li.
\newblock Deep imbalanced learning for multimodal emotion recognition in
  conversations.
\newblock \emph{IEEE Transactions on Artificial Intelligence}, 5\penalty0
  (12):\penalty0 6472--6487, 2024{\natexlab{b}}.

\bibitem[Michel et~al.(2023)Michel, Nikolentzos, Lutzeyer, and
  Vazirgiannis]{michel2023path}
Gaspard Michel, Giannis Nikolentzos, Johannes~F Lutzeyer, and Michalis
  Vazirgiannis.
\newblock Path neural networks: Expressive and accurate graph neural networks.
\newblock In \emph{International Conference on Machine Learning}, pp.\
  24737--24755. PMLR, 2023.

\bibitem[Mohri(2018)]{mohri2018foundations}
Mehryar Mohri.
\newblock Foundations of machine learning, 2018.

\bibitem[Peng et~al.(2024)Peng, Chen, Shou, and Chen]{peng2024carat}
Cheng Peng, Ke~Chen, Lidan Shou, and Gang Chen.
\newblock Carat: contrastive feature reconstruction and aggregation for
  multi-modal multi-label emotion recognition.
\newblock In \emph{Proceedings of the AAAI Conference on Artificial
  Intelligence}, volume~38, pp.\  14581--14589, 2024.

\bibitem[Poria et~al.(2017)Poria, Cambria, Hazarika, Majumder, Zadeh, and
  Morency]{poria2017context}
Soujanya Poria, Erik Cambria, Devamanyu Hazarika, Navonil Majumder, Amir Zadeh,
  and Louis-Philippe Morency.
\newblock Context-dependent sentiment analysis in user-generated videos.
\newblock In \emph{Proceedings of the 55th annual meeting of the association
  for computational linguistics (volume 1: Long papers)}, pp.\  873--883, 2017.

\bibitem[Poria et~al.(2018)Poria, Hazarika, Majumder, Naik, Cambria, and
  Mihalcea]{poria2018meld}
Soujanya Poria, Devamanyu Hazarika, Navonil Majumder, Gautam Naik, Erik
  Cambria, and Rada Mihalcea.
\newblock Meld: A multimodal multi-party dataset for emotion recognition in
  conversations.
\newblock \emph{arXiv preprint arXiv:1810.02508}, 2018.

\bibitem[Qin et~al.(2025)Qin, Liu, Tang, Liu, Wang, Huang, and
  Zhang]{qin2025mental}
Jinghui Qin, Changsong Liu, Tianchi Tang, Dahuang Liu, Minghao Wang, Qianying
  Huang, and Rumin Zhang.
\newblock Mental-perceiver: Audio-textual multi-modal learning for estimating
  mental disorders.
\newblock In \emph{Proceedings of the AAAI Conference on Artificial
  Intelligence}, volume~39, pp.\  25029--25037, 2025.

\bibitem[Qiu et~al.(2020)Qiu, Chen, Dong, Zhang, Yang, Ding, Wang, and
  Tang]{qiu2020gcc}
Jiezhong Qiu, Qibin Chen, Yuxiao Dong, Jing Zhang, Hongxia Yang, Ming Ding,
  Kuansan Wang, and Jie Tang.
\newblock Gcc: Graph contrastive coding for graph neural network pre-training.
\newblock In \emph{Proceedings of the 26th ACM SIGKDD international conference
  on knowledge discovery \& data mining}, pp.\  1150--1160, 2020.

\bibitem[Redko et~al.(2017)Redko, Habrard, and Sebban]{redko2017theoretical}
Ievgen Redko, Amaury Habrard, and Marc Sebban.
\newblock Theoretical analysis of domain adaptation with optimal transport.
\newblock In \emph{Machine Learning and Knowledge Discovery in Databases:
  European Conference, ECML PKDD 2017, Skopje, Macedonia, September 18--22,
  2017, Proceedings, Part II 10}, pp.\  737--753. Springer, 2017.

\bibitem[Ren et~al.(2021)Ren, Huang, Li, Song, and Nie]{ren2021lr}
Minjie Ren, Xiangdong Huang, Wenhui Li, Dan Song, and Weizhi Nie.
\newblock Lr-gcn: Latent relation-aware graph convolutional network for
  conversational emotion recognition.
\newblock \emph{IEEE Transactions on Multimedia}, 24:\penalty0 4422--4432,
  2021.

\bibitem[Shen et~al.(2018)Shen, Qu, Zhang, and Yu]{shen2018wasserstein}
Jian Shen, Yanru Qu, Weinan Zhang, and Yong Yu.
\newblock Wasserstein distance guided representation learning for domain
  adaptation.
\newblock In \emph{Proceedings of the AAAI conference on artificial
  intelligence}, volume~32, 2018.

\bibitem[Shou et~al.(2022)Shou, Meng, Ai, Yang, and Li]{shou2022conversational}
Yuntao Shou, Tao Meng, Wei Ai, Sihan Yang, and Keqin Li.
\newblock Conversational emotion recognition studies based on graph
  convolutional neural networks and a dependent syntactic analysis.
\newblock \emph{Neurocomputing}, 501:\penalty0 629--639, 2022.

\bibitem[Shou et~al.(2023)Shou, Ai, Meng, Zhang, and Li]{shou2023graphunet}
YunTao Shou, Wei Ai, Tao Meng, FuChen Zhang, and KeQin Li.
\newblock Graphunet: Graph make strong encoders for remote sensing
  segmentation.
\newblock In \emph{2023 IEEE 29th International Conference on Parallel and
  Distributed Systems (ICPADS)}, pp.\  2734--2737. IEEE, 2023.

\bibitem[Shou et~al.(2024{\natexlab{a}})Shou, Ai, Du, Meng, Liu, and
  Yin]{shou2024efficient}
Yuntao Shou, Wei Ai, Jiayi Du, Tao Meng, Haiyan Liu, and Nan Yin.
\newblock Efficient long-distance latent relation-aware graph neural network
  for multi-modal emotion recognition in conversations.
\newblock \emph{arXiv preprint arXiv:2407.00119}, 2024{\natexlab{a}}.

\bibitem[Shou et~al.(2024{\natexlab{b}})Shou, Liu, Cao, Meng, and
  Dong]{shou2024low}
Yuntao Shou, Huan Liu, Xiangyong Cao, Deyu Meng, and Bo~Dong.
\newblock A low-rank matching attention based cross-modal feature fusion method
  for conversational emotion recognition.
\newblock \emph{IEEE Transactions on Affective Computing}, 16\penalty0
  (2):\penalty0 1177--1189, 2024{\natexlab{b}}.

\bibitem[Shou et~al.(2024{\natexlab{c}})Shou, Meng, Ai, Zhang, Yin, and
  Li]{shou2024adversarial}
Yuntao Shou, Tao Meng, Wei Ai, Fuchen Zhang, Nan Yin, and Keqin Li.
\newblock Adversarial alignment and graph fusion via information bottleneck for
  multimodal emotion recognition in conversations.
\newblock \emph{Information Fusion}, 112:\penalty0 102590, 2024{\natexlab{c}}.

\bibitem[Shou et~al.(2024{\natexlab{d}})Shou, Yan, Yuan, Cao, Zhao, and
  Meng]{shou2024graph}
Yuntao Shou, Peiqiang Yan, Xingjian Yuan, Xiangyong Cao, Qian Zhao, and Deyu
  Meng.
\newblock Graph domain adaptation with dual-branch encoder and two-level
  alignment for whole slide image-based survival prediction.
\newblock \emph{arXiv preprint arXiv:2411.14001}, 2024{\natexlab{d}}.

\bibitem[Shou et~al.(2025{\natexlab{a}})Shou, Cao, Liu, and
  Meng]{shou2025masked}
Yuntao Shou, Xiangyong Cao, Huan Liu, and Deyu Meng.
\newblock Masked contrastive graph representation learning for age estimation.
\newblock \emph{Pattern Recognition}, 158:\penalty0 110974, 2025{\natexlab{a}}.

\bibitem[Shou et~al.(2025{\natexlab{b}})Shou, Cao, and Meng]{shou2025spegcl}
Yuntao Shou, Xiangyong Cao, and Deyu Meng.
\newblock Spegcl: Self-supervised graph spectrum contrastive learning without
  positive samples.
\newblock \emph{IEEE Transactions on Neural Networks and Learning Systems},
  2025{\natexlab{b}}.

\bibitem[Shou et~al.(2025{\natexlab{c}})Shou, Cao, Yan, Hui, Zhao, and
  Meng]{shou2025graph}
Yuntao Shou, Xiangyong Cao, Peiqiang Yan, Qiao Hui, Qian Zhao, and Deyu Meng.
\newblock Graph domain adaptation with dual-branch encoder and two-level
  alignment for whole slide image-based survival prediction.
\newblock In \emph{Proceedings of the IEEE/CVF International Conference on
  Computer Vision}, pp.\  19925--19935, 2025{\natexlab{c}}.

\bibitem[Shou et~al.(2025{\natexlab{d}})Shou, Lan, and
  Cao]{shou2025contrastive}
Yuntao Shou, Haozhi Lan, and Xiangyong Cao.
\newblock Contrastive graph representation learning with adversarial cross-view
  reconstruction and information bottleneck.
\newblock \emph{Neural Networks}, 184:\penalty0 107094, 2025{\natexlab{d}}.

\bibitem[Shou et~al.(2025{\natexlab{e}})Shou, Meng, Ai, and
  Li]{shou2025dynamic}
Yuntao Shou, Tao Meng, Wei Ai, and Keqin Li.
\newblock Dynamic graph neural ode network for multi-modal emotion recognition
  in conversation.
\newblock In \emph{Proceedings of the 31st International Conference on
  Computational Linguistics}, pp.\  256--268, 2025{\natexlab{e}}.

\bibitem[Shou et~al.(2025{\natexlab{f}})Shou, Meng, Ai, and
  Li]{shou2025multimodal}
Yuntao Shou, Tao Meng, Wei Ai, and Keqin Li.
\newblock Multimodal large language models meet multimodal emotion recognition
  and reasoning: A survey.
\newblock \emph{arXiv preprint arXiv:2509.24322}, 2025{\natexlab{f}}.

\bibitem[Shou et~al.(2025{\natexlab{g}})Shou, Meng, Ai, and
  Li]{shou2025revisiting}
Yuntao Shou, Tao Meng, Wei Ai, and Keqin Li.
\newblock Revisiting multi-modal emotion learning with broad state space models
  and probability-guidance fusion.
\newblock In \emph{Joint European Conference on Machine Learning and Knowledge
  Discovery in Databases}, pp.\  509--525. Springer, 2025{\natexlab{g}}.

\bibitem[Shou et~al.(2025{\natexlab{h}})Shou, Meng, Ai, Yin, and
  Li]{shou2025cilf}
Yuntao Shou, Tao Meng, Wei Ai, Nan Yin, and Keqin Li.
\newblock Cilf-ciae: Clip-driven image--language fusion for correcting inverse
  age estimation.
\newblock \emph{Neural Networks}, pp.\  108518, 2025{\natexlab{h}}.

\bibitem[Shou et~al.(2025{\natexlab{i}})Shou, Yao, Meng, Ai, Chen, and
  Li]{shou2025gsdnet}
Yuntao Shou, Jun Yao, Tao Meng, Wei Ai, Cen Chen, and Keqin Li.
\newblock Gsdnet: Revisiting incomplete multimodality-diffusion emotion
  recognition from the perspective of graph spectrum.
\newblock In \emph{Proceedings of the Thirty-Fourth International Joint
  Conference on Artificial Intelligence, IJCAI-25. International Joint
  Conferences on Artificial Intelligence Organization}, pp.\  6182--6190,
  2025{\natexlab{i}}.

\bibitem[Shou et~al.(2026{\natexlab{a}})Shou, Ai, Meng, and Li]{shou2026graph}
Yuntao Shou, Wei Ai, Tao Meng, and Keqin Li.
\newblock Graph diffusion models: A comprehensive survey of methods and
  applications.
\newblock \emph{Computer Science Review}, 59:\penalty0 100854,
  2026{\natexlab{a}}.

\bibitem[Shou et~al.(2026{\natexlab{b}})Shou, Meng, Ai, Fu, Yin, and
  Li]{shou2026comprehensive}
Yuntao Shou, Tao Meng, Wei Ai, Fangze Fu, Nan Yin, and Keqin Li.
\newblock A comprehensive survey on multi-modal conversational emotion
  recognition with deep learning.
\newblock \emph{ACM Transactions on Information Systems}, 44\penalty0
  (2):\penalty0 1--48, 2026{\natexlab{b}}.

\bibitem[Sun et~al.(2026)Sun, Zhang, Hong, Zhu, and Gao]{sun2026boomda}
Jun Sun, Xinxin Zhang, Simin Hong, Jian Zhu, and Xiang Gao.
\newblock Boomda: Balanced multi-objective optimization for multimodal domain
  adaptation.
\newblock In \emph{Proceedings of the AAAI Conference on Artificial
  Intelligence}, volume~40, pp.\  25700--25708, 2026.

\bibitem[Sun et~al.(2022)Sun, Zhou, He, Wang, and Wang]{sun2022gppt}
Mingchen Sun, Kaixiong Zhou, Xin He, Ying Wang, and Xin Wang.
\newblock Gppt: Graph pre-training and prompt tuning to generalize graph neural
  networks.
\newblock In \emph{Proceedings of the 28th ACM SIGKDD Conference on Knowledge
  Discovery and Data Mining}, pp.\  1717--1727, 2022.

\bibitem[Sun et~al.(2024)Sun, Wei, Ni, Liu, Song, Wang, and Nie]{sun2024muti}
Teng Sun, Yinwei Wei, Juntong Ni, Zixin Liu, Xuemeng Song, Yaowei Wang, and
  Liqiang Nie.
\newblock Muti-modal emotion recognition via hierarchical knowledge
  distillation.
\newblock \emph{IEEE Transactions on Multimedia}, 2024.

\bibitem[Tang et~al.(2025)Tang, Pan, Zheng, Zhou, Sui, Zhu, Deng, and
  Kuai]{tang2025pose}
Bin Tang, Ke-Qi Pan, Miao Zheng, Ning Zhou, Jia-Lu Sui, Dandan Zhu, Cheng-Long
  Deng, and Shu-Guang Kuai.
\newblock Pose as a modality: A psychology-inspired network for personality
  recognition with a new multimodal dataset.
\newblock In \emph{Proceedings of the AAAI Conference on Artificial
  Intelligence}, volume~39, pp.\  1538--1546, 2025.

\bibitem[Tao et~al.(2025)Tao, Li, Zang, and Gao]{tao2025multi}
Chuanqi Tao, Jiaming Li, Tianzi Zang, and Peng Gao.
\newblock A multi-focus-driven multi-branch network for robust multimodal
  sentiment analysis.
\newblock In \emph{Proceedings of the AAAI Conference on Artificial
  Intelligence}, volume~39, pp.\  1547--1555, 2025.

\bibitem[Tellamekala et~al.(2023)Tellamekala, Amiriparian, Schuller, Andr{\'e},
  Giesbrecht, and Valstar]{tellamekala2023cold}
Mani~Kumar Tellamekala, Shahin Amiriparian, Bj{\"o}rn~W Schuller, Elisabeth
  Andr{\'e}, Timo Giesbrecht, and Michel Valstar.
\newblock Cold fusion: Calibrated and ordinal latent distribution fusion for
  uncertainty-aware multimodal emotion recognition.
\newblock \emph{IEEE Transactions on Pattern Analysis and Machine
  Intelligence}, 46\penalty0 (2):\penalty0 805--822, 2023.

\bibitem[Tu et~al.(2024)Tu, Xie, Liang, Wang, and Xu]{tu2024adaptive}
Geng Tu, Tian Xie, Bin Liang, Hongpeng Wang, and Ruifeng Xu.
\newblock Adaptive graph learning for multimodal conversational emotion
  detection.
\newblock In \emph{Proceedings of the AAAI Conference on Artificial
  Intelligence}, volume~38, pp.\  19089--19097, 2024.

\bibitem[Villani et~al.(2009)]{villani2009optimal}
C{\'e}dric Villani et~al.
\newblock \emph{Optimal transport: old and new}, volume 338.
\newblock Springer, 2009.

\bibitem[Wagner et~al.(2023)Wagner, Triantafyllopoulos, Wierstorf, Schmitt,
  Burkhardt, Eyben, and Schuller]{wagner2023dawn}
Johannes Wagner, Andreas Triantafyllopoulos, Hagen Wierstorf, Maximilian
  Schmitt, Felix Burkhardt, Florian Eyben, and Bj{\"o}rn~W Schuller.
\newblock Dawn of the transformer era in speech emotion recognition: closing
  the valence gap.
\newblock \emph{IEEE Transactions on Pattern Analysis and Machine
  Intelligence}, 45\penalty0 (9):\penalty0 10745--10759, 2023.

\bibitem[Wang et~al.(2024{\natexlab{a}})Wang, Zhang, Liu, Wu, Hu, Yu, and
  Wang]{10680310}
Ye~Wang, Wei Zhang, Ke~Liu, Wei Wu, Feng Hu, Hong Yu, and Guoyin Wang.
\newblock Dynamic emotion-dependent network with relational subgraph
  interaction for multimodal emotion recognition.
\newblock \emph{IEEE Transactions on Affective Computing}, pp.\  1--14,
  2024{\natexlab{a}}.
\newblock \doi{10.1109/TAFFC.2024.3461148}.

\bibitem[Wang et~al.(2024{\natexlab{b}})Wang, Wang, Liu, and
  Yin]{wang2024degree}
Yingxu Wang, Mengzhu Wang, Siwei Liu, and Nan Yin.
\newblock Degree distribution based spiking graph networks for domain
  adaptation.
\newblock \emph{arXiv preprint arXiv:2410.06883}, 2024{\natexlab{b}}.

\bibitem[Wen et~al.(2024)Wen, Cao, Shen, Yang, Liu, and
  Sun]{wen2024personality}
Zhiyuan Wen, Jiannong Cao, Jiaxing Shen, Ruosong Yang, Shuaiqi Liu, and Maosong
  Sun.
\newblock Personality-affected emotion generation in dialog systems.
\newblock \emph{ACM Transactions on Information Systems}, 42\penalty0
  (5):\penalty0 1--27, 2024.

\bibitem[Wu et~al.(2020)Wu, Pan, Zhou, Chang, and Zhu]{wu2020unsupervised}
Man Wu, Shirui Pan, Chuan Zhou, Xiaojun Chang, and Xingquan Zhu.
\newblock Unsupervised domain adaptive graph convolutional networks.
\newblock In \emph{Proceedings of the Web Conference 2020}, pp.\  1457--1467,
  2020.

\bibitem[Xu et~al.(2025)Xu, Yuan, Wei, Wu, Wang, and Wu]{xu2025multiple}
Qinfu Xu, Shaozu Yuan, Yiwei Wei, Jie Wu, Leiquan Wang, and Chunlei Wu.
\newblock Multiple feature refining network for visual emotion distribution
  learning.
\newblock In \emph{Proceedings of the AAAI Conference on Artificial
  Intelligence}, volume~39, pp.\  8924--8932, 2025.

\bibitem[Yang et~al.(2020)Yang, Deng, Liu, and Tao]{yang2020heterogeneous}
Xu~Yang, Cheng Deng, Tongliang Liu, and Dacheng Tao.
\newblock Heterogeneous graph attention network for unsupervised
  multiple-target domain adaptation.
\newblock \emph{IEEE Transactions on Pattern Analysis and Machine
  Intelligence}, 44\penalty0 (4):\penalty0 1992--2003, 2020.

\bibitem[Yang et~al.(2024)Yang, Li, Cheng, Zhang, and Wang]{yang2024emotion}
Zhenyu Yang, Xiaoyang Li, Yuhu Cheng, Tong Zhang, and Xuesong Wang.
\newblock Emotion recognition in conversation based on a dynamic complementary
  graph convolutional network.
\newblock \emph{IEEE Transactions on Affective Computing}, 2024.

\bibitem[Yin et~al.(2019)Yin, Kannan, and Bartlett]{yin2019rademacher}
Dong Yin, Ramchandran Kannan, and Peter Bartlett.
\newblock Rademacher complexity for adversarially robust generalization.
\newblock In \emph{International conference on machine learning}, pp.\
  7085--7094. PMLR, 2019.

\bibitem[Yin et~al.(2022)Yin, Shen, Li, Wang, Luo, Chen, Luo, and
  Hua]{yin2022deal}
Nan Yin, Li~Shen, Baopu Li, Mengzhu Wang, Xiao Luo, Chong Chen, Zhigang Luo,
  and Xian-Sheng Hua.
\newblock Deal: An unsupervised domain adaptive framework for graph-level
  classification.
\newblock In \emph{Proceedings of the 30th ACM International Conference on
  Multimedia}, pp.\  3470--3479, 2022.

\bibitem[Yin et~al.(2023)Yin, Shen, Wang, Luo, Luo, and Tao]{10113198}
Nan Yin, Li~Shen, Mengzhu Wang, Xiao Luo, Zhigang Luo, and Dacheng Tao.
\newblock Omg: Towards effective graph classification against label noise.
\newblock \emph{IEEE Transactions on Knowledge and Data Engineering},
  35\penalty0 (12):\penalty0 12873--12886, 2023.

\bibitem[Yin et~al.(2024{\natexlab{a}})Yin, Shen, Chen, Hua, and
  Luo]{10.1145/3687468}
Nan Yin, Li~Shen, Chong Chen, Xian-Sheng Hua, and Xiao Luo.
\newblock Sport: A subgraph perspective on graph classification with label
  noise.
\newblock 18\penalty0 (9), 2024{\natexlab{a}}.
\newblock ISSN 1556-4681.
\newblock URL \url{https://doi.org/10.1145/3687468}.

\bibitem[Yin et~al.(2024{\natexlab{b}})Yin, Yu, Lin, Liu, Sonke, and
  Gavves]{yindomain}
Wenzhe Yin, Shujian Yu, Yicong Lin, Jie Liu, Jan-Jakob Sonke, and Stratis
  Gavves.
\newblock Domain adaptation with cauchy-schwarz divergence.
\newblock In \emph{The 40th Conference on Uncertainty in Artificial
  Intelligence}, 2024{\natexlab{b}}.

\bibitem[You et~al.(2023)You, Chen, Wang, and Shen]{you2023graph}
Yuning You, Tianlong Chen, Zhangyang Wang, and Yang Shen.
\newblock Graph domain adaptation via theory-grounded spectral regularization.
\newblock In \emph{The Eleventh International Conference on Learning
  Representations}, 2023.

\bibitem[Young et~al.(2018)Young, Cambria, Chaturvedi, Zhou, Biswas, and
  Huang]{young2018augmenting}
Tom Young, Erik Cambria, Iti Chaturvedi, Hao Zhou, Subham Biswas, and Minlie
  Huang.
\newblock Augmenting end-to-end dialogue systems with commonsense knowledge.
\newblock In \emph{Proceedings of the AAAI conference on artificial
  intelligence}, volume~32, 2018.

\bibitem[Yuan et~al.(2023)Yuan, Huang, Li, Yuan, Pun, and Zhong]{yuan2023rba}
Lin Yuan, Guoheng Huang, Fenghuan Li, Xiaochen Yuan, Chi-Man Pun, and Guo
  Zhong.
\newblock Rba-gcn: Relational bilevel aggregation graph convolutional network
  for emotion recognition.
\newblock \emph{IEEE/ACM Transactions on Audio, Speech, and Language
  Processing}, 31:\penalty0 2325--2337, 2023.

\bibitem[Zhang et~al.(2025)Zhang, Huang, Xu, Fan, Xiao, Dai, and
  Huang]{zhang2025core}
Bowen Zhang, Zhichao Huang, Guangning Xu, Xiaomao Fan, Mingyan Xiao, Genan Dai,
  and Hu~Huang.
\newblock Core knowledge learning framework for graph.
\newblock In \emph{Proceedings of the AAAI Conference on Artificial
  Intelligence}, volume~39, pp.\  13179--13187, 2025.

\bibitem[Zhang et~al.(2024{\natexlab{a}})Zhang, Jin, Xu, Li, Xu, Wei, Liu, and
  Liu]{zhang2024camel}
Linhao Zhang, Li~Jin, Guangluan Xu, Xiaoyu Li, Cai Xu, Kaiwen Wei, Nayu Liu,
  and Haonan Liu.
\newblock Camel: capturing metaphorical alignment with context disentangling
  for multimodal emotion recognition.
\newblock In \emph{Proceedings of the AAAI Conference on Artificial
  Intelligence}, volume~38, pp.\  9341--9349, 2024{\natexlab{a}}.

\bibitem[Zhang et~al.(2024{\natexlab{b}})Zhang, Sun, Hong, and
  Li]{zhang2024amanda}
Xinxin Zhang, Jun Sun, Simin Hong, and Taihao Li.
\newblock Amanda: Adaptively modality-balanced domain adaptation for multimodal
  emotion recognition.
\newblock In \emph{Findings of the Association for Computational Linguistics
  ACL 2024}, pp.\  14448--14458, 2024{\natexlab{b}}.

\bibitem[Zhang et~al.(2024{\natexlab{c}})Zhang, Liu, Wang, Chen, Li, Bu, and
  He]{zhang2024collaborate}
Zhen Zhang, Meihan Liu, Anhui Wang, Hongyang Chen, Zhao Li, Jiajun Bu, and
  Bingsheng He.
\newblock Collaborate to adapt: Source-free graph domain adaptation via
  bi-directional adaptation.
\newblock In \emph{Proceedings of the ACM Web Conference 2024}, pp.\  664--675,
  2024{\natexlab{c}}.

\bibitem[Zhao et~al.(2023)Zhao, Wang, Shen, Xu, and Zhang]{zhao2023tdfnet}
Zhengdao Zhao, Yuhua Wang, Guang Shen, Yuezhu Xu, and Jiayuan Zhang.
\newblock Tdfnet: Transformer-based deep-scale fusion network for multimodal
  emotion recognition.
\newblock \emph{IEEE/ACM Transactions on Audio, Speech, and Language
  Processing}, 31:\penalty0 3771--3782, 2023.

\end{thebibliography}
